\documentclass[a4paper,11pt]{article}
\usepackage{jcappub} 
\usepackage{lineno}
\usepackage[dvipsnames]{xcolor}
\usepackage{xfrac}
\usepackage[T1]{fontenc} 
\usepackage{aas_macros}
\usepackage{natbib}
\usepackage{tabularx}
\usepackage{bm}
\usepackage{hyperref}
\usepackage{subcaption}
\usepackage{mathrsfs}
\usepackage{siunitx}
\usepackage{xspace}
\usepackage{xcolor}
\usepackage{soul}
\usepackage{colortbl}
\usepackage{threeparttable}
\usepackage{enumitem}
\usepackage{tikz}
\usetikzlibrary{positioning,arrows.meta,fit,backgrounds,calc}

\definecolor{inputfill}{HTML}{D6E4F2}
\definecolor{inputborder}{HTML}{5B84B1}
\definecolor{whitefill}{HTML}{FFFFFF}
\definecolor{grayfill}{HTML}{F6F7F9}
\definecolor{grayborder}{HTML}{C7CCD4}
\definecolor{yellowfill}{HTML}{FBEFCE}
\definecolor{yellowborder}{HTML}{D9A93A}
\definecolor{purplefill}{HTML}{EBE4F3}
\definecolor{purpleborder}{HTML}{7E57A6}
\definecolor{loopborder}{HTML}{6A3FA0}
\definecolor{greenfill}{HTML}{E3F0E4}
\definecolor{greenborder}{HTML}{5B9E5B}
\definecolor{greendark}{HTML}{2F6B3A}
\definecolor{labelgray}{HTML}{5A6472}
\definecolor{arrowcol}{HTML}{7B8794}

\newcommand{\lbl}[2]{{\bfseries\tiny\textsc{\textcolor{#1}{#2}}}}
\newcommand{\ttl}[1]{{\normalsize\bfseries #1}}
\newcommand{\bdy}[1]{{\footnotesize #1}}

\newcommand{\cell}[3]{\begin{minipage}[t][#1][t]{#2}\sffamily\raggedright #3\end{minipage}}
\newlist{blist}{itemize}{1}
\setlist[blist]{leftmargin=1.15em, labelsep=0.4em, topsep=2pt, itemsep=1.5pt,
                parsep=0pt, label=\raisebox{0.15ex}{\scriptsize\textbullet}}
\newcommand{\bitem}[1]{{\footnotesize\begin{blist}#1\end{blist}}}
\def\TW{3.2}

\def\B4R{15.851}     
    
\def\FIGC{9.9755}    
\tikzset{
  box/.style={draw, rounded corners=5pt, line width=0.9pt, inner sep=5pt,
              anchor=north west},
  arr/.style={-{Stealth[length=6pt,width=6pt]}, line width=1.1pt, draw=arrowcol},
}
\newcommand{\figscale}{0.76}

\definecolor{lightgray}{gray}{0.9} 

\newcommand{\EM}[1]{\ensuremath{\mathrm{EM}\!\left(#1\right)}}

\newcommand{\SB}{\ensuremath{S\left(\theta\right)}}
\newcommand{\CR}{\ensuremath{\mathrm{CR}}\xspace}
\newcommand{\lx}{\ensuremath{L_X}\xspace}

\newcommand{\CM}{CHEX-MATE\xspace}
\newcommand{\ls}{L38\xspace}
\newcommand{\MSZ}{\ensuremath{M_{500c}^{Y_{\mathrm SZ}}}}
\newcommand{\RSZ}{\ensuremath{R_{500c}^{Y_{\mathrm SZ}}}}
\newcommand{\bSZ}{\ensuremath{b_{{\mathrm SZ}}}}
\newcommand{\MWL}{\ensuremath{M_{500c}}}
\newcommand{\ms}{\ensuremath{M_\odot}}

\newcommand{\efeds}{eFEDs\xspace}

\arxivnumber{1234.56789} 

\title{Baryonification IV: Constraining baryonic feedback with X-ray gas fractions}
\author[a]{Jozef Bucko,}
\author[b]{Andrina Nicola,}
\author[c]{Aurel Schneider,}
\author[b]{Michael Kova\v{c},}
\author[d,e]{Sambit K. Giri,}
\author[f]{Robert Reischke,}
\author[g,h]{Esra Bulbul,}
\author[i]{Nicolas Clerc,}
\author[j,k]{Ang Liu,}
\author[l,m]{Iacopo Bartalucci,}
\author[a]{Alexandre Refregier}

\affiliation[a]{Institute for Particle Physics and Astrophysics, ETH Zurich, Wolfgang Pauli Strasse 27, 8093 Zurich, Switzerland}
\affiliation[b]{Jodrell Bank Centre for Astrophysics, Department of Physics and Astronomy, The University of Manchester, Manchester M13 9PL, UK}
\affiliation[c]{Department of Astrophysics, University of Zurich, Winterthurerstrasse 190, 8057 Zurich, Switzerland}
\affiliation[d]{Department of Astronomy and Oskar Klein Centre, AlbaNova, Stockholm University, SE-10691 Stockholm, Sweden}
\affiliation[e]{Van Swinderen Institute for Particle Physics and Gravity, University of Groningen, Nijenborgh 4, 9747 AG Groningen, The Netherlands}
\affiliation[f]{Argelander-Institut für Astronomie, Universität Bonn, Auf dem Hügel 71, D-53121 Bonn, Germany}
\affiliation[g]{Max Planck Institute for Extraterrestrial Physics, Giessenbachstrasse 1, 85748 Garching, Germany}
\affiliation[h]{Faculty of Physics, Fakultät für Physik, Ludwig-Maximilians-Universität, Scheinerstr. 1, 81679 München, Germany}
\affiliation[i]{Univ Toulouse, CNES, CNRS, IRAP, Toulouse, France}
\affiliation[j]{School of Physics and Astronomy, Beijing Normal University, Beijing
100875, China}
\affiliation[k]{Institute for Frontiers in Astronomy and Astrophysics, Beijing Normal
University, Beijing 102206, China}
\affiliation[l]{Università degli studi di Roma ‘Tor Vergata’, Via della ricerca scientifica, 1, 00133 Roma, Italy}
\affiliation[m]{INAF – Istituto di Astrofisica Spaziale e Fisica Cosmica di Milano, via A. Corti 14, I-20133 Milano, Italy}

\emailAdd{jbucko@phys.ethz.ch}

\abstract{ 
Baryonic feedback redistributes gas around dark matter halos, suppressing the matter power spectrum at scales now probed by weak lensing surveys. X-ray observations directly trace this hot gas, and are one of the main probes of its distribution and properties. We present a forward-modelling framework, built on the baryonification model, linking the three-dimensional gas density and temperature profiles of groups and clusters to observed X-ray surface brightness and luminosity profiles on one side, and to matter power spectrum suppression on the other. We validate the model against independent three-dimensional density reconstructions from the literature, and examine our temperature and metallicity treatment in the group-scale regime. Applying this framework to the SZ-selected CHEX-MATE and X-ray-selected eFEDs samples, we measure gas fractions across the group-to-cluster mass range while accounting for X-ray selection effects, with the first published gas fractions based on CHEX-MATE data. Combining both samples, we derive a joint constraint on the hot gas fraction retained by groups and clusters as a function of mass and on the baryonic suppression of the matter power spectrum. We find $f_{\rm gas} = 0.029 \pm 0.006$ at $M_{500c} = 3\times 10^{13}M_\odot$, $f_{\rm gas} = 0.078 \pm 0.004$ at $M_{500c} = 3\times 10^{14}M_\odot$, and suppression of 6\% at $k=1\,h/\rm Mpc$ and 23\% at $k=5\,h/\rm Mpc$. Our findings are consistent with recent kinematic Sunyaev-Zel'dovich results, hinting at strong feedback. We also show that the $L_X$--$M$ relation is degenerate with feedback strength, and that different feedback scenarios produce distinct X-ray profile shapes that map onto the same $L_X$--$M$ point. This work is a first step towards extending the framework to forward-model diffuse X-ray emission at the map level for simulation-based inference in upcoming wide-area X-ray surveys such as eROSITA.
}

\begin{document}
\maketitle
\flushbottom

\section{Introduction}
\label{sec:intro}

X-ray observations encode rich information about the thermodynamic state of galaxy clusters and groups, and particularly about the distribution of hot gas. In the past, surveys such as \textit{Chandra} and \textit{XMM-Newton} collected deep observations of massive clusters ($M\gtrsim10^{14} M_\odot$) to study in detail the dynamical state of these objects \cite{Vikhlinin_chandra1,Vikhlinin_chandra2, Pratt_xmm1,Eckert_xmm2,Arnaud_2002,Arnaud_2010_xmm,McDonald_2017_chandra}. However, apart from thermodynamic properties of clusters (such as temperature, pressure, or entropy profiles), X-ray observations provide a promising insight into the effects of baryonic feedback on the large-scale structure of the universe \cite{cosmoOWLS,bahamas,flamingo,frontiere,Schneider_2020,Schneider_2022,Kovač_2025,Ferreira_Xray_WL,laposta2024}. 

By measuring the flux of X-ray photons within a given energy band, one can derive the X-ray surface brightness as a function of angular radius. Under the assumption of fully ionised gas, these surface brightness profiles are proportional to the projected square of the gas number density. Various techniques have been developed to deproject the resulting two-dimensional profiles and reconstruct the three-dimensional gas density distribution, most of which assume spherical symmetry \cite{Kriss_83_deprojection, Sanders_18_mbproj2}, although non-spherical deprojection methods have also been explored \cite{zaroubi_axial_deprojection}. Once cluster redshifts are known, angular radii can be converted into physical scales within a chosen cosmological model. If the total cluster mass within a given radius is available, the gas mass fraction can be compared to the cosmic baryon fraction, quantifying how efficiently active galactic nuclei (AGN) feedback at a given mass scale ejects gas from the clusters. Alternatively, the suppression of the total matter power spectrum due to baryons can serve as a useful indicator of the strength of baryonic feedback, which can be, in principle, obtained if surface brightness profiles together with the total masses of clusters at different masses are measured.

Gas fractions have been used in the past to both obtain constraints on feedback \cite{Grandis_2024,Kovač_2025,Bigwood_2024,siegel_2025}, as well as to calibrate subgrid models of multiple hydrodynamical simulations \cite{cosmoOWLS,bahamas,flamingo}. However, as several studies argue, inferring feedback strength from gas fractions can be challenging, particularly because of uncertainty in cluster total mass. Several methods have been developed to estimate cluster masses, such as assuming hydrostatic equilibrium \cite{ettori_2013_hse}, using scaling relations between cluster mass and Compton-$y$ \cite{Arnaud2010,planck_scaling_relations}, investigating kinematic properties of satellite galaxies (so-called dynamical mass estimates) \cite{Sereno_2025} or using weak gravitational lensing to estimate the total mass of clusters \cite{hoekstra_wl_masses}. In particular, the first two methods are well known to underestimate total mass, with misestimation ranging from 10-40\% \cite{ettori_2013_hse,Planck_XX_szbias,Planck_XXIV_szbias, linden_sz_bias, sereno_ettori_2017_sz_bias}. Some recent dynamical mass estimates agree well with estimates based on weak lensing \cite{Sereno_2025}, which, by construction, are sensitive to total mass and therefore subject to less severe biases. However, because weak lensing profiles are rarely available for X-ray clusters, many past studies relied on biased mass estimates and applied a mass-bias parameter, usually defined as the ratio of biased to unbiased mass for an object. The uncertainty in mass bias (i.e. in total mass) can significantly alter the gas fraction measurements, as it affects the total gas mass of the cluster via the characteristic (virial) radius of the cluster (which depends on the total mass as $M^{1/3}$), and has the total mass directly in the denominator. In addition, the gas fraction value is reported at a given total halo mass, making this quantity even more volatile. All of this makes it challenging to derive reliable constraints on the strength of baryonic feedback from gas fractions. Moreover, calibrating feedback models on gas fractions ensures a correct gas content of clusters at a given mass; however, it does not impose any requirement on a specific redistribution of gas inside the virial radius. Therefore, quantities such as stacked X-ray or Sunyaev-Zeldovich profiles are not uniquely defined from such a calibration. On the other hand, feedback calibration or baryonic feedback constraints obtained directly from X-ray profiles, unlike those using gas fractions, largely alleviate these limitations.

Another substantial systematic effect in X-ray observations is sample selection. Selecting clusters based on their X-ray emission may lead to biases, as very bright clusters are over-represented in such a sample \cite{pratt_2009_selection_bias,Vikhlinin_chandra2}. As a result, biases at the level of $\mathcal{O}(10\%)$ up to a factor of two in the mass-bolometric luminosity relation have been reported  \cite{stanek_2x_sel_fct_bias,Viklinin_chandra_sel_fct_biases,sel_fct_eff_87percent}. A different approach is selection based on Compton-$y$ (often referred to as SZ-selection), which has been shown to be a robust, although biased, mass proxy \cite{nagai06_SZ}. SZ selection benefits from less dramatic scaling of the signal with electron density, however, it has been shown to over-represent hot mergers \cite{SZ_selection_hot_mergers}. Optically selected samples have also been studied, with detection relying on optical richness instead of X-ray emission. Although less prone to the Malmquist bias typical of X-selection, projection effects, miscentering, or richness-mass scatter still need to be handled correctly \cite{Popesso_2024} to avoid sample contamination. Understanding selection effects in X-ray samples is therefore a necessary step to derive unbiased results from these measurements, for example, on the strength of baryonic feedback. This is a non-trivial challenge for theoretical models, as selection functions are typically built using complex simulations \cite{clerc_erass_selection_function}. However, these potential systematics in X-ray analyses can be more effectively controlled in simulation-based forward modelling of X-ray observations.

In this work, we develop a modelling framework to derive the constraints on baryonic feedback from measured X-ray profiles of \efeds and \CM samples. First, we validate our model against measured data from the above samples. Next, and for the first time, we measure the gas fractions in the \CM sample in 4 mass-redshift bins. We perform a similar exercise and measure the gas fractions in 8 bins of the \efeds sample, carefully accounting for selection effects. Furthermore, we fit the low-redshift measured data points with a single BFC model to investigate the mass dependence of baryonic feedback. We use our results to compute the matter power spectrum suppression due to baryons and compare it with other results from the literature. We conclude with remarks on how the shapes of the X-ray profiles affect the general picture of baryonic feedback. This work is an initial but important step towards robust modelling of the X-ray data. In a follow-up paper, we aim to fit the X-ray profiles directly and compare to the current gas fraction analysis.

This work is organised as follows: in Sec.~\ref{sec:data} we describe the \CM and \efeds data used to perform our analysis; in Sec.~\ref{sec:modelling} we introduce the modelling of the stacked surface brightness profiles; Sec.~\ref{subsec:data_vector_and_covariance} provides details on how we build the data vectors and covariance matrices used for the final analysis; and
Sec.~\ref{sec:inference} discusses the aspects related to parameter inference. In Sec.~\ref{sec:validation}, we provide a set of validation and robustness tests, highlight the main results
in Sec.~\ref{sec:results}
and conclude in Sec.~\ref{sec:conclusion}.

\section{Observational data}
\label{sec:data}
The following section presents the observational data sets used in this analysis.
\subsection{CHEX-MATE}

The Cluster HEritage project with XMM-Newton -- Mass Assembly and Thermodynamics at the Endpoint of structure formation (CHEX-MATE; \citep{CHEX-MATE}) provides a sample of 116 galaxy clusters from the Planck catalogue \citep{Planck2016} observed in the X-ray band. Selected to have a signal-to-noise ratio $S/N > 6.5$ in the Planck PSZ2 catalogue, these clusters are intended to form a representative sample of the underlying cluster population at a given mass and redshift. This approach, while facing its own challenges such as hot-merger over-representation \cite{SZ_selection_hot_mergers}, overcomes the limitations of X-ray-selected samples, which are subject to selection biases that favour brighter-than-average objects \citep{CHEX-MATE}.

The CHEX-MATE cluster sample results from a two-tier selection process combining low-redshift systems and the most massive clusters of the PSZ2 catalogue. The clusters span redshifts $z=[0.05$--$0.6]$ and cover a mass range of $\MSZ=[1$--$13]\times10^{14}\,M_\odot$. The median mass of the sample is $\MSZ = 7.23 \times 10^{14}\,M_\odot$ and the median redshift is $z = 0.183$. The mass proxy \MSZ{} is derived by the Planck collaboration using the MMF3 SZ detection algorithm described in \cite{Planck2016}. This method measures the $\text{SZ}$ flux, proportional to the projected electron pressure of each cluster, and converts it into \MSZ{} via the \MSZ--$Y_\text{SZ}$ scaling relation calibrated in \cite{Arnaud2010}, under the assumption of self-similar evolution.

These masses are not corrected for hydrostatic mass bias \citep[section~5.3]{Planck2016} and are therefore biased low compared to masses obtained via weak lensing or dynamical analyses. The authors of \cite{Sereno_2025} derive dynamical mass estimates for the CHEX-MATE clusters and show them to be unbiased with respect to the masses inferred from weak lensing. Under the assumption that the bias does not depend on redshift or cluster luminosity, the authors find $b_\text{SZ} = -0.38 \pm 0.04$, with
\begin{equation}
\label{eq:bSZ}
M_{500c} = \frac{\MSZ}{1+b_\text{SZ}}.
\end{equation}
Recently, weak lensing masses for a subset of 41 \CM clusters were derived by \cite{chexmate_wl_masses}, yielding slightly lower SZ mass biases of $(1 + \bSZ) = 0.83 \pm 0.09$ assuming the mass and redshift dependence in the calibration and $(1 + \bSZ) = 0.72 \pm 0.11$, obtained respectively when assuming a fixed mass slope and when assuming no redshift evolution in the adopted regression model.

For every cluster in the CHEX-MATE sample, a measurement of the surface brightness profile, \SB, is available \cite{Bartalucci_2023}. The background signal contribution, originating from both the instrument and the sky, has been removed from the profiles, as described in \cite{Bartalucci_2023}. Consequently, our observational data set consists of 116 clusters with redshift, mass (from the PSZ2 catalogue), and surface brightness profile measurements.

At present, no gas fraction measurements or individual three-dimensional electron density profiles have been obtained for the CHEX-MATE clusters. However, \cite{Lyskova_2023} provide the average electron density profile for a subset of 38 CHEX-MATE clusters, which we refer to as \ls. The median SZ mass of this subsample is $\MSZ = 4.12 \times 10^{14}\,M_\odot$, and the median redshift is $z=0.11$.

\begin{figure}
    \centering
    \includegraphics[width=1.\linewidth]{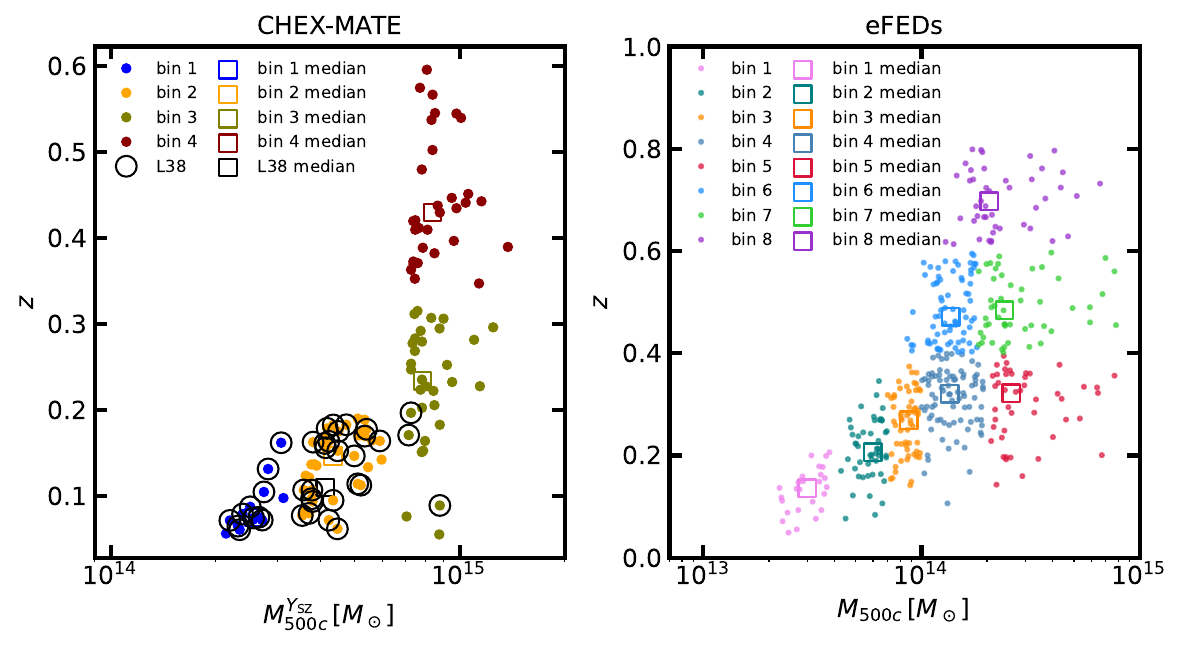}
    \caption{Mass and redshift distribution of the CHEX-MATE sample (left), highlighting different subsamples used throughout this work, and objects from the \efeds sample (right) used in our analysis. In both panels, circles represent the individual objects while squares mark the median mass and redshift of different data bins.}
    \label{fig:chm_bins}
\end{figure}

For the remainder of this work, we bin the CHEX-MATE clusters into subsamples, defining four subsets of the full sample, referred to as bin~1 through bin~4; their median masses, redshifts, and number of clusters are listed in Tab.~\ref{tab:chm_bins}. The mass and redshift distributions of these bins are shown in the left panel of Fig.~\ref{fig:chm_bins}. The coloured dots represent individual clusters, while the squares indicate the median mass and redshift of each bin. For the \ls\ subsample, which partially overlaps with bins~1, 2, and~3, we plot individual clusters using open circles instead of filled dots.

\begin{table}[]
    \renewcommand{\arraystretch}{1.2}
    \centering
    \setlength{\tabcolsep}{7pt}
    \begin{tabular}{l c c c c c}
    \hline\hline
        \CM & Size & \MSZ &
        $z$ &
        $T$ &
        $J(T,z)$ \\
         & &
        $[10^{14}\,M_\odot]$ & &
        $[\mathrm{keV}]$ &
        $[10^{-13}\,\mathrm{\,cm^{5}\,s^{-1}}]$ \\
        \hline
         bin 1       &  14 & 2.57 & 0.075 & 2.88 & 4.50 \\
         bin 2       &  39 & 4.34 & 0.147 & 4.14 & 3.92 \\
         bin 3       &  32 & 7.83 & 0.234 & 6.47 & 3.34 \\
         bin 4       &  31 & 8.36 & 0.430 & 7.41 & 2.66 \\
         \ls         &  38 & 4.12 & 0.110 & 4.04 & 4.14 \\
        \hline
    \end{tabular}
    \caption{Size, median SZ mass, median redshift, median
      temperature $T$, and emissivity $J$ for the \CM\ subsamples used in this work.}
    \label{tab:chm_bins}
\end{table}

\begin{table}[h]
  \centering
  \setlength{\tabcolsep}{7pt}
  \begin{tabular}{crrcrr}
    \hline\hline
    \efeds & Size &
      $ M_{500\mathrm{c}}$ &
      $ z$ &
      $T$ &
      $\Lambda_c(T, z)$ \\
        & &
      $[10^{13}\,M_\odot]$ & &
      $[\mathrm{keV}]$ &
      $[10^{-24}\,\mathrm{erg\,cm^{3}\,s^{-1}}]$ \\
    \hline
    bin 1 & 26 &   3.0 & 0.138 & 0.72 & 10.74 \\
    bin 2 & 49 &   5.6 & 0.211 & 1.04 & 9.19 \\
    bin 3 & 58 &   8.4 & 0.263 & 1.32 & 7.41 \\
    bin 4 & 84 &  13.5 & 0.310 & 1.77 & 6.34 \\
    bin 5 & 46 &  27.7 & 0.318 & 2.81 & 6.06 \\
    bin 6 & 73 &  13.3 & 0.470 & 1.65 & 6.53 \\
    bin 7 & 51 &  25.1 & 0.483 & 2.49 & 6.09 \\
    bin 8 & 45 &  22.5 & 0.697 & 2.20 & 6.13 \\
    \hline
  \end{tabular}
    \caption{Properties of the \efeds bins used in this analysis.
    'Size' refers to the number of objects, $ M_{500\mathrm{c}}$ the
    weighted-median mass,
    $z$ the weighted-median redshift, $T$ the
    temperature from the Chiu et~al.\ (2022) $T_\mathrm{X}$--$M$ scaling relation, and $\Lambda$ the 0.5--2\,keV cooling function
    ($Z=0.30\,Z_\odot$) at $(T,z)$.}
  \label{tab:efeds_bins}
\end{table}

\subsection{\efeds}

The {eROSITA} Final Equatorial-Depth Survey (\efeds) is a contiguous $\sim 140\,\mathrm{deg}^2$ X-ray field observed during the performance-verification phase of the {eROSITA} mission, designed to validate cluster detection, characterisation, and selection for the all-sky survey \cite{efeds_catalog}. The cluster catalogue contains 542 X-ray-selected (eROSITA) systems spanning a redshift range $0 \lesssim z \lesssim 1.3$, with a median redshift of $\sim 0.35$ and typical masses $M_{500c} \sim 10^{13} - 10^{15}\,M_\odot$. The cluster detection was performed in the soft band (0.5--2.0 keV) with a flux threshold of $\sim 10^{-14}\rm\,erg/s/cm^2$ \cite{efeds_catalog}. Photometric redshifts of the \efeds objects are available; furthermore, 470/542 objects were optically confirmed using imaging data from Hyper Suprime-Cam Subaru Strategic Program and from the Legacy Survey \cite{efeds_optical_confirmation}, and spectroscopic redshifts were derived \cite{efeds_spectroscopy}. Weak lensing mass calibration was carried out using shear measurements from Hyper Suprime-Cam Subaru Strategic Program (HSC-SSP) imaging, with tangential shear profiles of 313 \efeds clusters available. The final $M_{\rm 500c}$ masses were obtained through joint modelling of the count-rate and shear profiles of the clusters and groups \cite{efeds_mass_calibration}. The resulting weak lensing mass--mass--redshift ($M_{\rm WL}-M_{\rm 500c}-z$) scaling relation is used to convert the weak lensing mass measurement to the $M_{500c}$ masses. For each object in the sample, the rest-frame cumulative luminosity profile in the soft band is available across up to two orders of magnitude in projected radius. These profiles were derived via the X-ray image-level fitting procedure described in \cite{bahar_efeds_profiles}. Specifically, the aforementioned study modelled the two-dimensional surface brightness distribution by projecting a parametric Vikhlinin electron density profile \cite{Vikhlinin_chandra1}, convolving it with the eROSITA PSF and exposure map, and fitting it to the observed images via a Poisson likelihood, yielding density profiles under the assumption of spherical symmetry. As a result, the final luminosity profiles are not noisy observations but rather model-dependent best-fit quantities. X-ray selection is well known to introduce selection biases, especially near the flux limit. Furthermore, the same study publishes the best-fit Vikhlinin parameters for each object as well as the corresponding gas fraction measurements. The selection of \efeds clusters has been extensively studied \cite{Comparat+19_efeds_sel_fct, comparat+20_efeds_sel_fct,Liu+21_selection_fct} and is accounted for throughout this work (see Sec.~\ref{subsec:data_vector_and_covariance} for more details).

In the right panel of Fig.~\ref{fig:chm_bins}, we show the mass-redshift distribution of objects within the \efeds catalogue. We construct five low-redshift ($z<0.4$) bins, spaced roughly equally in log-mass, to trace the mass dependence of feedback at low redshift, while splitting the remaining clusters above $z=0.4$ into three bins, with a boundary at $z=0.6$. Note that we exclude objects below $\MWL<2\times 10^{13}\ms$, as these objects (with average temperatures below $\sim 1\,$keV) are challenging from both an observational and a modelling perspective, with details discussed in Sec.~\ref{subsubsec:modelling_LX_profiles}. We further exclude clusters above $z>0.8$. Note that the $x$-axis shows the weak lensing-calibrated masses, unlike the left panel, which shows the Planck (SZ) masses. For each bin, we show the median mass and redshift as an empty square. More details about the \efeds sample as well as the individual bins are summarised in Tab.~\ref{tab:efeds_bins}.

\section{Modelling}
\label{sec:modelling}

In this section, we describe our approach to modelling X-ray stacked profiles of massive clusters and groups, based on the BFC model. We also describe how this model connects to key predictors of baryonic feedback strength.

\subsection{Stacked X-ray profiles}
\label{subsec:modelling_stacked_profiles}

The X-ray profiles available for both the \CM and \efeds samples represent 2D-projected X-ray emission as a function of projected radius, albeit in slightly different formats. While the \CM sample provides surface brightness profiles (the count rate of X-ray photons per unit solid angle), the \efeds profiles represent rest-frame luminosity profiles in the soft band. Although these two formats are in principle directly linkable, we describe in separate sections below how we model the surface brightness profiles of \CM and the luminosity profiles of \efeds.

\subsubsection{Surface brightness profiles}
\label{subsubsec:modelling_CR_profiles}

Following \cite[][and references therein]{laposta2024}, we define the count rate of X-ray photons, \CR, sourced by the diffuse gas in the galaxy cluster at redshift $z$ and angular position $\hat{\boldsymbol{n}}$ on the sky as

\begin{equation}
    \CR(\hat{\boldsymbol{n}})=\int\frac{\mathrm{d}\chi}{4\pi(1+z(\chi))^3}\,n_{e}(\chi\hat{\boldsymbol{n}})\,n_H(\chi\hat{\boldsymbol{n}})J(T,Z,z(\chi)),
    \label{eq:CR_Int}
\end{equation}
where $\chi$ is the comoving distance, and $n_e$ and $n_H$ are the physical electron and hydrogen number densities, respectively. Throughout this work, we use the X-ray surface brightness \SB{} and count rate \CR interchangeably, such that $\SB = \CR(\theta)$. As can be seen, the galaxy cluster surface brightness profile depends on the radial distribution of electrons and hydrogen as well as on $J(T,Z,z(\chi))$, which we define as the instrument-convolved X-ray cooling function. Focusing first on the latter, this quantity depends on the temperature $T$ and metallicity $Z$ of the hot ICM in a cluster at redshift $z$, and is given by

\begin{equation}
    J(T, Z, z) \equiv \int \mathrm{d}E_o \, \phi(E_o) A(E_o) \Lambda_c((1 + z) E_o, T, Z) e^{-\sigma(E_o)N_{\rm HI}}, 
    \label{eq:J}
\end{equation}
where $\Lambda_c$ is the cooling function, $e^{-\sigma(E_o) N_{\rm HI}}$ is a transmission term that suppresses the X-ray signal due to the neutral hydrogen column density $N_{\rm HI}$ of the galaxy at the angular position of the cluster, with $\sigma(E_o)$ the photoabsorption cross-section, which depends on the energy $E_o$. In addition, $A(E_o)$ denotes the effective area of the detector, and the instrument bandpass $\phi (E_o)$ is defined as 
\begin{align}
\label{eq:instrument_bandpass}
    \phi (E_o)\equiv \int_{\widetilde{E}_{o,\text{min}}}^{\widetilde{E}_{o,\text{max}}} \mathrm{d}\widetilde{E}_o~\mathcal{M}(\widetilde{E}_o|E_o).
\end{align}
The quantity $\mathcal{M}(\widetilde{E}_o|E_o)$ is the energy redistribution matrix of the instrument, which relates the true, observer-frame energy of the photons hitting the detector, $E_o$, to the measured energy, $\widetilde{E}_o$. The observer-frame energy is related to the energy of the emitted photons as $E_e = (1+z)E_o$, and finally $\widetilde{E}_{o, \rm min}$ and $\widetilde{E}_{o, \rm max}$ are the edges of the observed band. To model $J(T, Z, z)$, we compute the cooling function $\Lambda_c$ using the Astrophysics Plasma Emission Code (APEC \cite{Smith_2001}) as implemented in the PYATOMDB library \citep{Foster_2012}, and we include the XMM-Newton energy redistribution and camera effective area specific to the CHEX-MATE observations. The observed energy band used to measure the surface brightness of the CHEX-MATE clusters is $\widetilde{E}_{o, \rm min}=0.7$~keV and $\widetilde{E}_{o, \rm max}=1.2$~keV \cite{Bartalucci_2023}. To compute the photoabsorption cross-section $\sigma(E_o)$, we use the phabs model within the XSPEC library \cite{xspec}, and we adopt the hydrogen column density measurements of \cite{bourdin_NHI_values} (see column 4 of Tab.~A.1).

\begin{figure}[h]
    \centering
    \includegraphics[width = \textwidth]{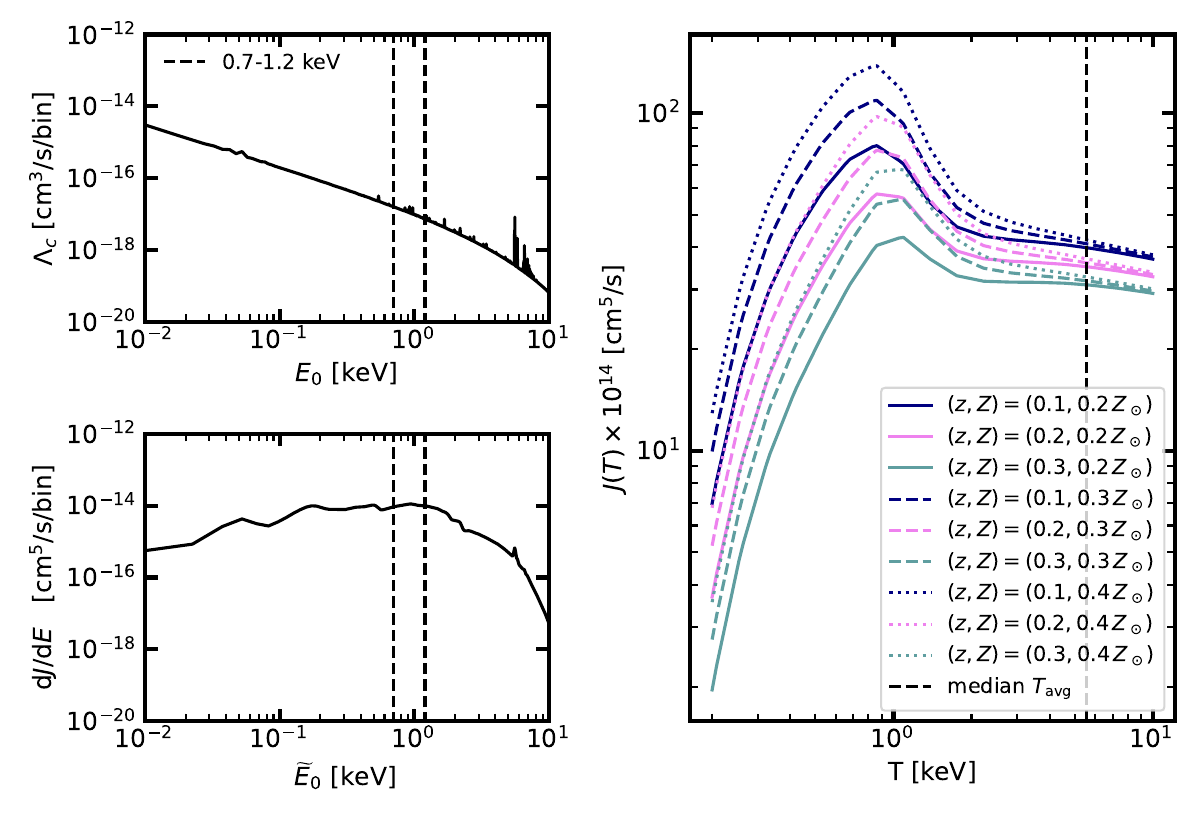}
    \caption{X-ray emission from hot plasma. The top left panel shows the cooling function $\Lambda_c$ as a function of the observer-frame energy $E_o$, while the bottom left panel shows the total emissivity $J$ in narrow observed energy bins $\widetilde{E}_o$, after convolution with the instrument response. The vertical dashed black lines indicate the CHEX-MATE energy band of 0.7--1.2~keV. The dependence of the total emissivity $J$ on redshift $z$ and metallicity $Z$ is shown in the right panel. Different colours correspond to redshifts of 0.1, 0.2, and 0.3, while different line styles represent metallicities of 0.2, 0.3, and 0.4, respectively.}
    \label{fig:emissivity}
\end{figure}

In the top left panel of Fig.~\ref{fig:emissivity}, we show the X-ray emission spectrum in the observer's frame as a function of $E_o$ (identical to the emitter's-frame spectrum up to a factor of $1+z$), while the spectrum convolved with the detector's effective area and energy redistribution (as a function of observed energy $\widetilde{E}_o$) is shown in the bottom left panel. The dashed black lines depict the energy band (0.7--1.2 keV) used to measure the emission from the \CM clusters. The right panel of the same figure shows the instrument-convolved cooling function $J$ as a function of ICM temperature for different metallicities and at redshifts $z=0.1,0.2,0.3$. Following Ref.~\cite{Bartalucci_2023}, we assume the plasma metallicity to be 25\% of the solar metallicity, $Z = 0.25 Z_\odot$, throughout this work. The mean temperature of the ICM in CHEX-MATE clusters ($T_{\rm avg} \simeq 5.6$~keV) is shown as a dashed black line in the same panel. As can be seen, the instrument-convolved cooling function $J$ depends only weakly on the plasma temperature for temperatures larger than $T \simeq 1$~keV. Therefore, some studies (e.g., \citep{Lyskova_2023, Bartalucci_2023}) neglect the radial dependence of cluster temperature and assume a constant average temperature when computing $J$.
Following \cite{Bartalucci_2023}, the average ICM temperature can be approximated as  
\begin{equation}
\label{eq:Tavg}
    T_{\rm avg} = 0.8 \times T_{500c} = 0.8 \times \frac{\mu_T m_p G \MSZ}{2 \RSZ},
\end{equation}
where $\mu_T$ is the mean molecular weight of the ICM, $G$ is the gravitational constant, and the factor of 0.8 represents the average value of the universal temperature profile derived in \cite{Ghirardini_2019} with respect to $T_{500c}$. The virial mass \MSZ{} in Eq.~\eqref{eq:Tavg} is the mass obtained from the mass-Compton $y$ ($Y_{\rm SZ}$) relation at the virial radius \RSZ{}. This mass definition of galaxy clusters has been shown to be biased with respect to the total gravitating mass inside the virial radius inferred, for example, from weak lensing or dynamical mass measurements \cite[][and references therein]{Sereno_2025}. We discuss our treatment of the mass bias in Sec.~\ref{sec:data}, and test and discuss the temperature approximation in more detail in Sec.~\ref{sec:validation}.

\subsubsection{Luminosity profiles}
\label{subsubsec:modelling_LX_profiles}

\begin{figure}
    \centering
    \includegraphics[width=0.85\linewidth]{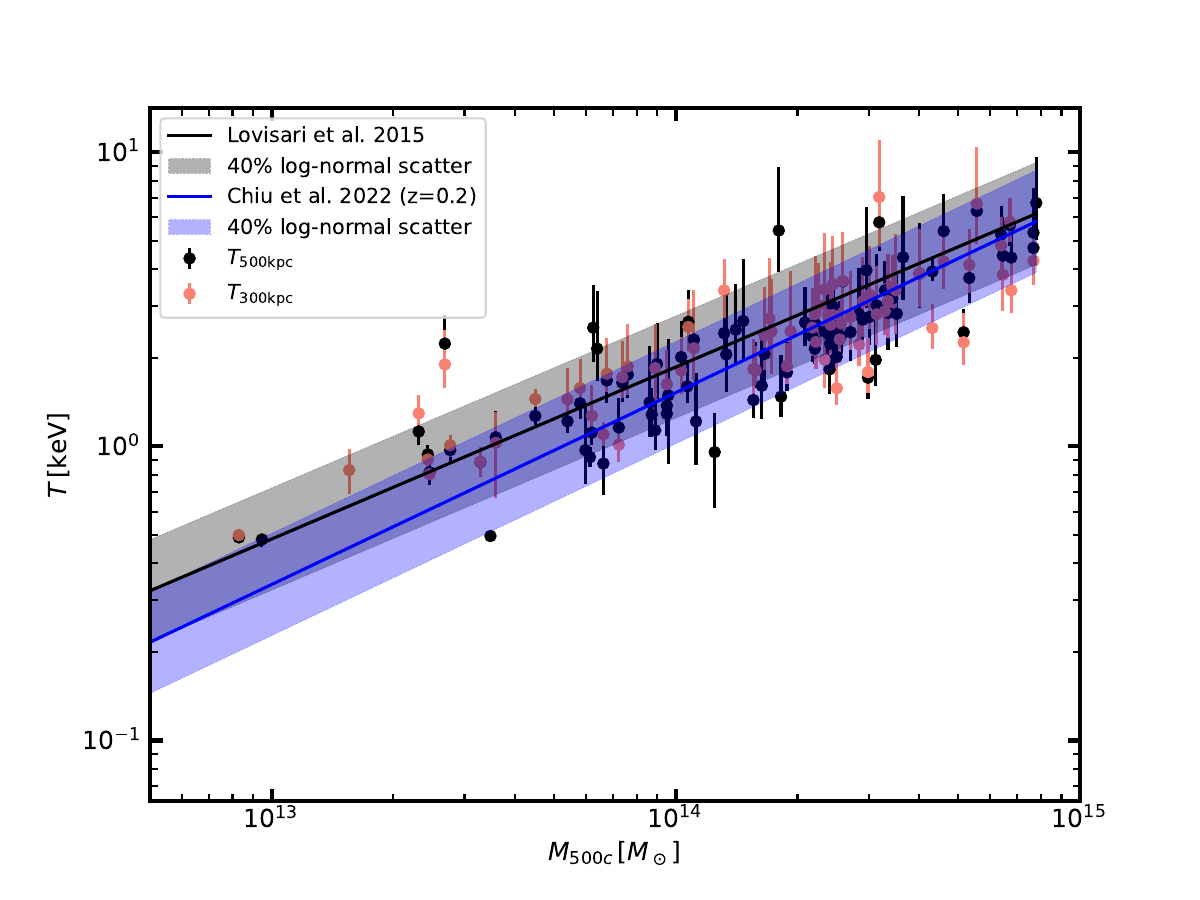}
    \caption{Temperature at a distance of 300~kpc (salmon) and 500~kpc (black) from the centre of 69 \efeds clusters as a function of the virial mass $M_{500c}$. The solid black line represents the scaling relation of \cite{Lovisari+15}, and the blue line is the one calibrated on the \efeds sample \cite{efeds_mass_calibration}. The shaded bands denote a 40\% log-normal scatter in temperature, assumed as a fiducial choice in our analysis.}
    \label{fig:T_vs_M_efeds}
\end{figure}

To model the rest-frame 2D-projected luminosity profile $L_X$ in the soft band, we can adopt Eq.~\eqref{eq:J} and remove the instrument term:
\begin{equation}
\label{eq:L_X}
    L_X(r_p) \equiv \int_{\rm los} {\rm d}l\, n_{e}(l\hat{\boldsymbol{n}})\,n_H(l\hat{\boldsymbol{n}}) \int_{\rm 0.5-2.0 keV} \mathrm{d}E \, \Lambda_c(E, T, Z),     
\end{equation}
with $r_p$ denoting the physical projected radius. In general, both the temperature $T$ and metallicity $Z$ depend on the distance from the cluster (or group) centre. Under the assumption of a constant temperature and metallicity per object, we can write
\begin{equation}
    L_X(r_p) \equiv {\rm EM}(r_p) \, \Lambda_{c, \rm soft}(T,Z).     
\end{equation}
Here, $L_X(r_p)$ is the product of the emission measure ${\rm EM}(r_{\rm p}) = \int n_e n_H \, dl$ and the cooling function integrated over the soft band, $\Lambda_{c, \rm soft}(T,Z) = \int_{\rm 0.5-2.0 keV} \mathrm{d}E \, \Lambda_c(E, T, Z)$. If, instead, a radial dependence in temperature or metallicity is assumed, we must treat the radius-dependent cooling function without the above simplification, i.e., by integrating Eq.~\eqref{eq:L_X} directly. The final modelled quantity, the cumulative luminosity profile $L_{X,\rm cum}$, can, in either scenario, be written as 
\begin{equation}
     L_{X,\rm cum}(r_p) = \int_0^{r_p}\, 2\pi r'_p \, L_X(r'_p)\, {\rm d}r'_p.    
     \label{eq:lx_cum}
\end{equation}
The temperatures in the \efeds sample are known only for a subsample of bright or close-by \efeds objects. They are measured at two different apertures, $r_p = \qty{300}{kpc}$ and $r_p=\qty{500}{kpc}$, for 102 objects in the sample (\cite{efeds_catalog}; objects with $>2\sigma$ temperature measurements within either 300~kpc or 500~kpc). We show these temperatures in Fig.~\ref{fig:T_vs_M_efeds} as salmon and black data points, respectively. The solid black line shows the temperature-mass scaling relation of \cite{Lovisari+15}, calibrated on XMM-Newton observations for a complete sample of galaxy groups selected based on the ROSAT All-Sky Survey, and the grey band marks the 40\% log-normal scatter around this scaling relation. In blue, we compare to the more recent result of \cite{efeds_mass_calibration}, calibrated directly on the \efeds sample. We model the temperature of the \efeds objects via the latter scaling relation, and choose the aforementioned 40\% log-normal scatter as an uncertainty on the average temperature in a given mass and redshift bin. Specifically, we introduce the departure from the $M-T$ scaling relation $dT_{\rm \efeds}$ and sample $T=T_{\rm M-T}\exp{dT_{\rm \efeds}}$, where $T_{M-T}$ stands for the temperature from the scaling relation. Finally, we impose a zero-mean Gaussian prior on the departure parameter with $\sigma_{dT_{\rm \efeds}} = 0.4$. As for the \CM sample, we test that assuming a constant temperature per object, instead of a temperature profile (which is not observationally available), is a good approximation; for more details, see App.~\ref{app:T_modelling}.

Modelling the temperatures and metallicities of X-ray emitting objects, especially for groups, poses a considerable challenge. While for objects in the \CM temperature range (5--6 keV) the temperature and metallicity dependence of the X-ray emission is minor (see Fig.~\ref{fig:emissivity}), for objects with temperatures of \qtyrange{1}{3}{\keV} the X-ray emission is not dominated by thermal bremsstrahlung but rather by line emission from metals; as a result, the X-ray emission varies significantly with both temperature and metallicity. For the latter, observational evidence across mass scales is rather sparse. For example, \cite{Lovisari+2019_metalicity} analysed metallicity profiles of 207 nearby ($z<0.1$) clusters observed by XMM-Newton and found that, although the metallicity can increase significantly close to the centre, the metallicity profiles become flat above \qtyrange{0.2}{0.3}{R_{500}}, with values typically around $Z\sim \qty{0.3}{Z_\odot}$. Using a different XMM-Newton sample, \cite{mernier+2017metal} derived the average abundance profiles of various metals and confirmed relatively large metallicity values (>0.6 $Z_\odot$) close to the cluster centres. On the other side, above $\sim 0.1\,R_{500}$, the abundance profiles, while not entirely flat, decrease from $Z\sim \qty{0.4}{Z_\odot}$ to $Z\sim \qty{0.2}{Z_\odot}$ close to the virial radius. Measurements of individual objects, such as \cite{Eckert+25_metal}, also show a dramatic increase in metallicity (up to $Z=Z_\odot$) below $r_p=\qty{100}{kpc}$ for the studied object before dropping to $\lesssim 0.2 Z_\odot$ at larger radii and becoming consistent with a flat profile. Although some previous works studying \efeds clusters \cite{Popesso_2024} assumed a fixed value of $Z=0.25Z_\odot$, we adopt the following strategy: first, we exclude radial scales below $r_p=\qty{100}{kpc}$ from all measured profiles. Furthermore, as suggested by the studies above, we assume a fixed metallicity value per object above $r_p=\qty{100}{kpc}$ and vary it within a Gaussian prior centred at $Z=\qty{0.3}{Z_\odot}$ with a standard deviation of 0.15, encompassing most of the observational evidence.

\subsection{Three dimensional profiles}
\label{subsec:modelling_density_profiles}

In addition to the cooling function and instrumental response, the second ingredient required to model X-ray surface brightness profiles is the three-dimensional gas density and temperature profiles. In this work, we model them using the so-called baryonification model \cite{Schneider_2015,Schneider_2019,arico_bacco_bf,schneider2025} and adopt the density profile parametrisation of \cite{schneider2025}. 

In the latest release of the model \cite{schneider2025}, the particles are further split into dark matter, gas, and stellar particles, providing a separate particle output for each component and enabling the modelling of observables such as the Sunyaev--Zeldovich signal or X-ray surface brightness. Specifically, we adopt the updated parametrisation of both the gas density and temperature profiles introduced in the latter work. In the following, we highlight the components of the adopted baryonification model most relevant to our analysis and refer the reader to the aforementioned papers for further details.

\subsubsection{Hot gas density profile and gas fractions}
The X-ray emission sourced by the ICM is directly connected to the physical density of the hot plasma inside the cluster (see Eq.~\ref{eq:CR_Int}), which, within the BFC model, is parameterised by the hot gas (`hga') density profile 
\begin{equation}
    \rho_{\text{hga}}(r) = \rho_{\text{hga,0}} \left[ 1 + \left( \frac{r}{r_c} \right)^\alpha\right]^{- \beta(M_{200c},M_c,\mu)/\alpha}  \left[ 1 + \left( \frac{r}{r_t} \right)^{\gamma} \right]^{-\frac{\delta}{\gamma}}.
    \label{eq:rho_hga_def}
\end{equation}
This is a three-slope power-law profile, which is constant below the core radius $r_c = \theta_c r_{200c}$, has slope $-\beta$ for intermediate radii, and slope $-\delta$ at large distances from the halo centre, well above the truncation radius $r_t = \varepsilon r_{200c}$. The transition between the three slopes is governed by the parameters $\alpha$ and $\gamma$, which we keep fixed at $1$ and $3/2$, respectively, in agreement with \cite{schneider2025}. 
The intermediate slope of the `hga' profile depends on the halo mass. This allows less massive halos to exhibit shallower gas profiles than high-mass halos, since AGN feedback expels gas more efficiently from their shallower gravitational potentials. The $\beta$ function is parameterised as:
\begin{equation}
    \beta(M_{200c},M_c,\mu) = \frac{3(M_{200c}/M_c)^{\mu}}{1 + (M_{200c}/M_c)^{\mu}},
    \label{eq:beta}
\end{equation}
which equals 3 at the highest masses and gradually decreases towards lower halo masses. The free model parameters $M_c$ and $\mu$ describe the mass scale where $\beta=3/2$ and the rate of this transition, respectively. Large values of $\mu$ produce a sharp transition from weak baryonic feedback in massive clusters ($M_{200c} > M_c$) to strong feedback in less massive clusters ($M_{200c} < M_c$), while smaller values of $\mu$ lead to a more continuous change in feedback strength across halos of different mass. The truncation radius $r_t$ depends on the halo mass and redshift via both the virial radius $r_{200c}$ and the parameter $\varepsilon = \varepsilon_0 - \nu\varepsilon_1$, where $\varepsilon_0 = 4$, $\varepsilon_1 = 0.5$, and $\nu$ is the peak height (see \cite{schneider2025} for more details). The normalisation factor $\rho_{\rm hga,0}$ is fixed such that the gas mass of a given halo is equal to the gas fraction times the total mass of a truncated NFW profile with the same $M_{200c}$. The gas density $\rho_{\rm hga}$ can then be connected to the number density of electrons $n_e$, hydrogen $n_H$, or the total gas number density $n_T$ via
\begin{equation}
    \label{eq:rho_gas_to_ni} n_i = \frac{\rho_{\rm gas}}{\mu_im_p},
\end{equation}
where $m_p$ is the mass of the proton and $\mu_i$ the mean molecular weight of species $i$, thus $\mu_e = \frac{2}{1+X_H}$, $\mu_H = \frac{1}{X_H}$, and $\mu_T = \frac{4}{3+5X_H}$, with hydrogen mass fraction $X_H=0.76$.

A widely used observational quantity representing the strength of baryonic feedback in galaxy clusters is the gas fraction, the ratio of gas mass to total cluster mass at the virial radius, typically reported at $r_{500c}$. Within the baryonification model, the gas fraction at a given halo mass $M_{500c}$ can be obtained as the ratio of the hot gas and total (`dmb') masses inside the virial radius $r_{500c}$:
\begin{equation}
\label{eq:fgas}
    f_{\mathrm{gas}, 500c} = \frac{M_{\rm hga}(r_{500c})}{M_{\rm dmb}(r_{500c})},
\end{equation}
where $M_i(r) = \int_0^r 4\pi r'^2 \rho_i(r') \rm{d}r'$ is the mass of species $i$ enclosed inside radius $r$. The total `dmb' density profile within the BFC model is defined as the sum of the gas, stellar, and dark matter profiles. For brevity, we do not introduce the detailed parametrizations of these profiles here, and instead refer the interested reader to \cite{schneider2025}.

In summary, we describe the three-dimensional gas density profile with a set of four free parameters, namely $M_c$, $\mu$, $\theta_{\rm co}$, and $\delta$, all defined at a given redshift $z$. In the case of data spanning different redshifts, we assume an explicit redshift dependence for the core radius, $\theta_{\rm co}(z) = \theta_{\rm co}(1+z)^{0.5}$.
This is motivated by the profiles from hydrodynamical simulations studied in \cite{schneider2025}.

\subsubsection{Temperature profile}

Modelling gas temperature and pressure — both at the level of radial profiles and full three-dimensional fields — is one of the key novelties of the baryonification approach introduced by \cite{schneider2025}. This framework enables the modelling of X-ray and Sunyaev--Zel'dovich (SZ) observables and, ultimately, their combination with weak lensing measurements. In this work, we aim to model X-ray emission, which, within the BFC framework, depends on the plasma temperature \(T\), derived from the thermal pressure \(P_{\rm th}\) via the ideal-gas law:
\begin{equation}
    P_{\rm th} = n_{\rm tot} \, k_B T ,
\end{equation}
where \(n_{\rm tot}\) is the particle number density of the gas. We assume hydrostatic equilibrium to relate the pressure gradient to the underlying mass distribution:
\begin{equation}
    \frac{\mathrm{d}P_{\rm tot}}{\mathrm{d}r} = -\rho_{\rm hga}(r)\frac{G M_{\rm dmb}(r)}{r^2}.
\end{equation}
The total pressure $P_{\rm tot}$ is modelled as the sum of thermal and non-thermal components, with the latter following the empirical prescription of \cite{Shaw_2010}:
\begin{equation}
    P_{\rm nth}(r) = P_{\rm tot}(r)\,\alpha_{\rm nth}(z) \left(\frac{r}{r_{200\rm c}}\right)^{n_{\rm nth}}.
\end{equation}
Consequently, the thermal pressure becomes
\begin{equation}
    P_{\rm th}(r) = P_{\rm tot}(r)\Big[1 - \alpha_{\rm nth}(z)\left(\frac{r}{r_{200\rm c}}\right)^{n_{\rm nth}}\Big].
\end{equation}
Finally, the temperature profile is obtained from the ideal-gas law:
\begin{equation}
    \label{eq:T_BFC}
    T(r) = \frac{m_p \, \mu_T \, P_{\rm th}(r)}{k_B \, \rho_{\rm hga}(r)}.
\end{equation}
Throughout this work, we adopt fixed values \(\alpha_{\rm nth} = 0.18\) and \(n_{\rm nth} = 0.8\) \citep{schneider2025, Kovač_2025}.

\subsection{Matter power spectrum}
\label{subsec:modelling_pk_emulator}

The suppression of the matter power spectrum with respect to a dark-matter-only universe provides a useful measure of the strength of baryonic feedback. In this work, we aim to translate the constraints on BFC parameters from X-ray observations to constraints on the suppression of the matter power spectrum. This suppression is a natural model prediction of the BFC model. However, baryonifying a simulation with sufficient resolution to yield a reliable power spectrum up to sufficiently high wavenumbers $k$ can be computationally expensive and inefficient for on-the-fly applications. To overcome this limitation, we build an emulator that efficiently predicts power-spectrum suppression.

For training, we use a set of PKDGRAV3 $N$-body simulations \cite{Potter_2016} with $L_{\rm box}=256~{\rm Mpc}/h$ and $N_{\rm part}=1024^3$. The halo catalogue is obtained with the AMIGA halo finder \cite{AHF}, considering only halos containing at least 200 dark matter particles. This corresponds to a minimum halo mass of $M_{h, \rm min}=2.5\times 10^{11}\,M_\odot/h$. We train the emulator across six redshifts between $z=0$ and $z=3$, and for three separate values of $q_2 = [0.5,0.7,0.9]$ (we linearly interpolate the emulator output for arbitrary $q_2 \in [0.5-0.9]$). The latter parameter controls the dark matter back-reaction to the gas expansion driven by feedback processes; see \cite{schneider2025} for more details. By comparing baryonified power spectra at the minimum and maximum redshifts for several simulation resolutions, we find that the resolution described above is required to capture the influence of smaller halos on the power spectrum, especially at higher redshift. In addition, we perform $N$-body runs for different realisations of the initial conditions and verify that the resulting power spectrum suppressions in the $k$-range relevant for baryonic feedback ($k\sim 0.05-10\,h$/Mpc) are unaffected by cosmic variance.
\begin{figure}
    \centering
    \includegraphics[width=1\linewidth]{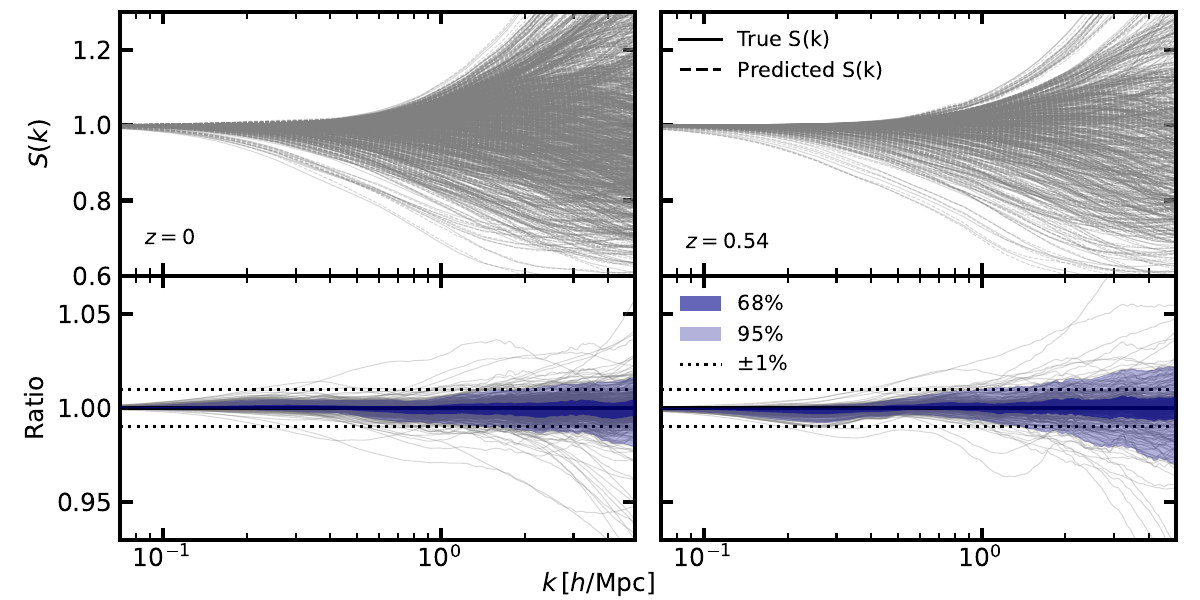}
    \caption{Performance of the power spectrum emulator on the validation set of BFC models at $z=0$ (left) and $z=0.54$ (right). The upper panels show the suppression of the matter power spectrum for different models within the prior range from baryonified $N$-body simulations (solid lines) and the emulated suppressions (dashed lines). The bottom panels show the ratios of the emulated and measured suppressions. The bands indicate the 68\% (dark blue) and 95\% (light blue) envelopes, respectively, and the dotted black lines in the bottom panels show the 1\% error budget.}
    \label{fig:emulator_pks}
\end{figure}
At each redshift, we run $\sim 2000$ baryonifications for $z \leq 1$ and $\sim 1000$ for $z > 1$, sampling different BFC parameter sets using Latin hypercube sampling in an eight-dimensional parameter space. Within the emulator, we vary $M_c, \mu, \theta_{\rm co}, \delta, \eta, \eta_\delta, N_{\rm star}$, and the baryon fraction $f_b$ (for the definition of these parameters, see \cite{schneider2025}). The ranges adopted for the BFC parameters during the emulator construction are summarised in Tab.~\ref{tab:emulator_priors}. Additionally, we fix the following parameter values in accordance with \cite{schneider2025,Kovač_2025}:
$c_{\rm iga}=0.1$, $\gamma=1.5$, $\varepsilon_1=0.5$, ${\rm halo}_{\rm excl}=0.4$, and $q_1=0.25$.
The cosmological parameters of the $N$-body simulation used to train the emulator are $\Omega_m=0.317$, $\Omega_b=0.049$, $\sigma_8=0.812$, $h=0.6727$, and $n_s=0.9649$. The initial conditions were set at $z_{\rm ini}=49$.

We train the emulator using a methodology similar to that described in \cite{Giri_2021}. The matter power-spectrum suppression data in the training set is first compressed using principal component analysis, retaining 20 principal components. A machine learning framework then learns a mapping from the 8 baryonic parameters to these retained components. We train a neural network (NN) with 4 hidden layers of 256 nodes each and ReLU activations, replacing the Gaussian Process Regression used in \cite{Giri_2021}. Hyperparameters (learning rate, weight decay) of this NN are tuned via Optuna \cite{optuna_2019}. The use of a NN makes this emulator lighter on memory, and it is publicly available through the \texttt{BCemu}\footnote{\url{https://github.com/sambit-giri/BCemu}} package. This emulator reach a precision of 1\% or better across all considered redshifts for $k\lesssim 5\, h/\rm Mpc$ at 68\% confidence. This is illustrated in Fig.~\ref{fig:emulator_pks}, where we show the performance of the emulator at redshift $z=0$ (left panel) and $z=0.54$ (right panel). The top panels show pairs of the true matter power-spectrum suppression $S(k)$ (measured from dark-matter-only and baryonified particle outputs of the $N$-body simulations, solid lines) and the emulated suppressions (dashed lines). In the bottom panels, we show the ratio of emulated and measured $S(k)$, with the dark blue band indicating the 68\% and the light blue band the 95\% confidence interval.

\begin{table}[h]
\renewcommand{\arraystretch}{1.2} 
\centering
\begin{tabular}{l c}
\hline
\hline
Parameter & Emulator Range \\
\hline
$\log_{10}M_c$       & [11, 15] \\
$\mu$                & [0, 3] \\
$\theta_{\rm co}$    & [0.01, 0.80] \\
$\delta$             & [2, 12] \\
$\eta$               & [-0.2, 0.2] \\
$\eta_\delta$        & [0.0, 0.4] \\
$N_{\rm star}$       & [0.00,0.05] \\
$f_b$                & [0.1,0.2] \\
\hline
\end{tabular}
\caption{Training domain of the BFC parameters varied in the power spectrum emulator. }
\label{tab:emulator_priors}
\end{table}

\section{Data vector and covariance matrix}
\label{subsec:data_vector_and_covariance}
In this section, we describe the process of building the binned data vectors from the individual per-object measured X-ray profiles.

\begin{table}[h]
  \centering
  \begin{threeparttable}
  \renewcommand{\arraystretch}{1.3}
  \setlength{\tabcolsep}{8pt}
  \begin{tabular}{l l l}
    \hline\hline
    Bin & Mass $[10^{14}\,M_\odot]$ & $z$ \\
    \hline
    \multicolumn{3}{l}{\CM — $M_\mathrm{SZ}$} \\[2pt]
    bin 1 & $< 3.4$               & all\tnote{*} \\
    bin 2 & $[3.4,\ 6.2)$        & all\tnote{*} \\
    bin 3 & $\geq 6.2$            & $< 0.33$ \\
    bin 4 & $\geq 6.2$            & $\geq 0.33$ \\
    \ls   & \multicolumn{2}{l}{38 morphologically-relaxed clusters \cite{Lyskova_2023}} \\
    \hline
    \multicolumn{3}{l}{\efeds — $M_{500c}$} \\[2pt]
    bin 1 & $[0.2,\ 0.4)$        & all\tnote{*} \\
    bin 2 & $[0.4,\ 0.7)$        & all\tnote{*} \\
    bin 3 & $[0.7,\ 1.0)$        & $< 0.4$ \\
    bin 4 & $[1.0,\ 2.0)$        & $< 0.4$ \\
    bin 5 & $\geq 2.0$            & $< 0.4$ \\
    bin 6 & $< 1.8$        & $[0.4,\ 0.6)$ \\
    bin 7 & $\geq 1.8$            & $[0.4,\ 0.6)$ \\
    bin 8 & $\geq 0.5$            & $[0.6,\ 0.8)$ \\
    \hline
  \end{tabular}
  \begin{tablenotes}
    \small
    \item[*] Objects only at low redshift, due to the flux cut.
  \end{tablenotes}
  \end{threeparttable}
  \caption{Bin definitions for the \CM\ and eFEDS samples.
    \CM\ bins are defined in terms of the Planck SZ mass \MSZ.
    eFEDS bins are defined in terms of the true cluster mass at $r_{500c}$ derived from weak lensing.}
  \label{tab:bin_definitions}
\end{table}

\subsection{\CM}
For \CM, we define four bins in order to study both the mass and redshift dependence of feedback using stacked profiles as shown in Tab.~\ref{tab:bin_definitions}.
We plot the mass-redshift properties of these bins in the left panel of Fig.~\ref{fig:chm_bins}. We assign the clusters to their respective bins based on measured redshift and SZ mass. For each bin, we compute the median redshift and mass, arriving at $\bar{z}=0.075, 0.147, 0.234$, and $0.430$, and $\overline{\MSZ}/10^{14}\ms = 2.57, 4.34, 7.83$, and $8.36$, respectively. We do not account for measurement errors on the median masses of the \CM sample, as the dominant source of mass uncertainty is the bias between the true and SZ masses introduced in Eq.~\eqref{eq:bSZ}, at the level of 10\%--50\% (see App.~\ref{app:bias_effects} for more detailed discussion about mass bias).

To build the data vector in each bin $i$, we first linearly interpolate each profile to 8 logarithmically spaced radial bins, i.e. 8 data points along the projected radius $r_p$. For \CM profiles, we consider a radial range between $r_{p,\rm min}=0.03\RSZ$ and $r_{p,\rm max}=0.88\RSZ$, the radial range available in all the measured profiles. The linear interpolation is not problematic, as the measured data are provided at much finer radial intervals than our final radial binning. Our data vector is defined as the median of all count rate profiles $\CR_i$ assigned to a given bin, evaluated on a common $r_p$ grid. The \CM sample is constructed to be minimally biased \cite{Bartalucci_2023}, and thus we do not account for any selection function in the \CM data.

For the cluster sample (and subsamples) considered in our work, the intrinsic population scatter dominates over the measurement uncertainties of individual clusters, and this scatter is correlated across neighbouring radial bins. Instead of incorporating this scatter into our model, we account for it as part of the covariance matrix. We construct a separate covariance matrix for each of the \CM bins.
To estimate the covariance of the median profile \(\CR_{i}\) in a given bin, we use bootstrap resampling of the cluster sample. We generate \(N_{\mathrm{boot}}\) bootstrap realisations by sampling clusters with replacement and compute the median profile \(\CR_{i}^{(b)}\) in each realisation \(b\). The bootstrap mean of the median profiles is defined as
\[
\overline{\CR}_{i} = \frac{1}{N_{\mathrm{boot}}} \sum_{b=1}^{N_{\mathrm{boot}}} \CR_{i}^{(b)} .
\]
The covariance matrix is then estimated as
\[
\mathrm{Cov}(i,j) = \frac{1}{N_{\mathrm{boot}} - 1} \sum_{b=1}^{N_{\mathrm{boot}}}
\left(\CR_{i}^{(b)} - \overline{\CR}_{i}\right)
\left(\CR_{j}^{(b)} - \overline{\CR}_{j}\right),
\]
where \(\CR_{i}^{(b)}\) denotes the median profile in radial bin \(i\) for bootstrap realisation \(b\). We find that the resulting covariance matrices converge for $N_{\rm boot}=10\,000$.
The final correlation matrices for \CM are shown in the four panels of Fig.~\ref{fig:covariances_bins1234}. At $r_p/\RSZ\sim 0.25$, the correlations between radial bins change qualitatively, with cross-correlations becoming smaller at larger radii. Clusters display larger intrinsic scatter near the core radius, while towards the outskirts the profiles approach a universal scaling. In the latter regime of larger radii, the shape of the surface brightness profile becomes largely insensitive to the physical processes (thus profile-to-profile scatter is suppressed), predominantly affecting the profiles close to the cores.

\begin{figure}
    \centering
\begin{tikzpicture}[x=1cm, y=1cm, scale=\figscale,
                    every node/.append style={transform shape}]
 
\def\th{4.3cm}
 
\node[box, fill=inputfill,  draw=inputborder] (b1) at (0.0,0) {\cell{\th}{\TW cm}{%
  \lbl{inputborder}{INPUT\newline }\\[3pt]
  \ttl{eFEDs catalogue}\\[1pt]
  \bdy{(542 objects)}\\[3pt]
  \bitem{%
    \item Cumulative $L_X$ profiles
    \item masses $M_{500c}$
    \item redshifts $z$}}};
 
\node[box, fill=whitefill, draw=grayborder] (b2) at (4.1,0) {\cell{\th}{\TW cm}{%
  \lbl{labelgray}{STEP 1: BINNING\newline}\\[3pt]
  \ttl{Mass--redshift bin assignment}\\[3pt]
  \bitem{%
    \item 8 bins: 5 low-$z$ + 3 high-$z$
    \item boundaries on $M_{500c}$
    \item 482 objects assigned}}};
 
\node[box, fill=yellowfill, draw=yellowborder] (b3) at (8.2,0) {\cell{\th}{\TW cm}{%
  \lbl{yellowborder}{STEP 2: SELECTION FUNCTION}\\[3pt]
  \ttl{Detection probability \& SF weights}\\[3pt]
  \scriptsize
     $p_i\! = \!p(L_X, z, T_{\mathrm{exp}}, EM_0) $\\ $\Rightarrow$
     $w_i = 1/p_i$}};
 
\node[box, fill=yellowfill, draw=yellowborder] (b4) at (12.3,0) {\cell{\th}{\TW cm}{%
  \lbl{yellowborder}{STEP 3: OUTLIER SCREENING}\\[3pt]
  \ttl{Forward-modelled weight distribution}\\[3pt]
  \bitem{%
    \item Large-area mock $\rightarrow$ 95th percentile of $\max(w_i)$
    \item flag \& remove outliers}}};
 
\node[box, fill=whitefill, draw=grayborder] (b5) at (16.4,0) {\cell{\th}{\TW cm}{%
  \lbl{labelgray}{STEP 4: STACKED PROFILE\newline}\\[3pt]
  \ttl{SF-weighted median stack}\\[3pt]
  \bitem{%
    \item $\hat{L}_X(r_p) = \mathrm{med}\{L_{X,i};\, w_i\}$
    \item $\bar{M}_{500c}$, $\bar{z}$ per bin
    \item 432 objects used}}};
 
\foreach \a/\b in {b1/b2,b2/b3,b3/b4,b4/b5}{\draw[arr] (\a.east) -- (\b.west);}
 
\draw[loopborder, dashed, line width=1.3pt, rounded corners=6pt]
      (2.05,-5.4) rectangle (17.901,-8.55);
\node[anchor=north west]
      at (2.33,-5.47)
      {\sffamily\bfseries\footnotesize\textcolor{loopborder}%
       {BOOTSTRAP LOOP\quad $B = 1 \ldots N_{\mathrm{boot}}$}};
 
\def\ph{1.9cm}
\node[box, fill=purplefill, draw=purpleborder] (b5a) at (2.65,-6.05) {\cell{\ph}{6.67cm}{%
  \lbl{purpleborder}{STEP 5A: RESAMPLE}\\[3pt]
  \ttl{SF-weighted resample with replacement}\\[5pt]
  \bdy{Draw catalogue with prob $\propto w_i$}}};
 
\node[box, fill=purplefill, draw=purpleborder] (b5b) at (10.2755,-6.05) {\cell{\ph}{6.67cm}{%
  \lbl{purpleborder}{STEP 5B: BIN \& STACK}\\[3pt]
  \ttl{Reassign bins $\rightarrow$ median mass \& profile}\\[6pt]
  \bdy{$M^{(b)}_{500c}\,,\ \ L^{(b)}_{X,\mathrm{cum}}(r_p)$}}};
 
\draw[arr] (b5a.east) -- (b5b.west);
 
\draw[arr] (\FIGC,-4.75) -- (\FIGC,-5.38);
 
\def\gh{2.3cm}
\node[box, fill=greenfill, draw=greenborder] (oa) at (0,-9.4) {\cell{\gh}{9.22cm}{%
  \lbl{greendark}{OUTPUT A: MASS PRIOR}\\[3pt]
  \ttl{Bootstrap mass variance}\\[6pt]
  \bdy{$\sigma^2_M = \mathrm{Var}_b\!\left[M^{(b)}\right]$}\\[4pt]
  \bdy{$b_{\mathrm{eFEDs}} \sim \mathcal{N}(0,\, \sigma^2_M)$}}};
 
\node[box, fill=greenfill, draw=greenborder] (ob) at (10.3755,-9.4) {\cell{\gh}{9.22cm}{%
  \lbl{greendark}{OUTPUT B: DATA VECTOR \& COVARIANCE}\\[3pt]
  \ttl{Bootstrap covariance matrix}\\[6pt]
  \bdy{$\mathbf{C} = \mathrm{Cov}_b\!\left[L^{(b)}_{X,\mathrm{cum}}\right]$}\\[4pt]
  \bdy{\textit{(cross-bin covariance neglected)}}}};
 
\draw[arr] (\FIGC,-8.65) -- (\FIGC,-9.3);
 
\end{tikzpicture}

    \caption{Overview of the \efeds data preprocessing pipeline.}
    \label{fig:efeds_pipeline}
\end{figure}
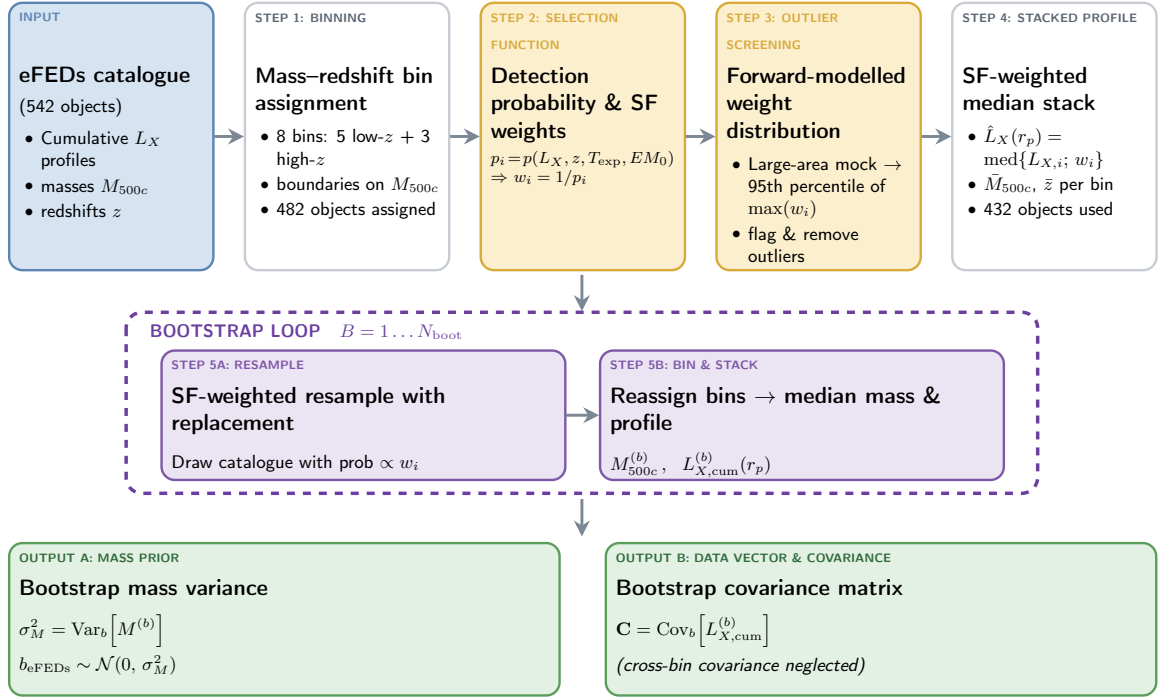

\subsection{\efeds}
\label{sec:efeds_datavector}

We construct the \efeds stacked data vector following the five-step
pipeline illustrated in Fig.~\ref{fig:efeds_pipeline}. Starting from
the full \efeds catalogue of 542 clusters and groups, we build a
selection function (SF)-weighted median $L_X$ profile in each of eight mass--redshift bins,
and estimate the associated covariance and mass uncertainties via
bootstrap resampling. The radial range of each profile is
$r_{p,\mathrm{min}} = 100\,\mathrm{kpc}$ to
$r_{p,\mathrm{max}} = r_{500c}$ (see
Sec.~\ref{subsec:modelling_stacked_profiles}).

\paragraph{Step 1: Mass--redshift bin assignment.}
We partition the catalogue into eight bins defined on object mass
$M_{500c}$ and redshift $z$ (Tab.~\ref{tab:bin_definitions},
Fig.~\ref{fig:chm_bins}): five low-$z$ bins ($z < 0.4$, bins~1--5) and
three high-$z$ bins ($z > 0.4$, bins~6--8). Bin boundaries are chosen
to balance the signal-to-noise ratio, number of objects per bin, and
coverage of the mass--redshift plane. Of the 542 catalogue objects,
482 fall within the union of these bins; the remaining 60 lie outside
all bin boundaries and are discarded.

\paragraph{Step 2: Selection-function weighting.}
Because \efeds is an X-ray flux-limited survey, bright sources are
over-represented relative to the underlying cluster population (Malmquist
bias). We correct for this by assigning each cluster $i$ a weight
\begin{equation}
  w_i = \frac{1}{p_i},
  \label{eq:sf_weight}
\end{equation}
where $p_i$ is the detection probability of that object. Crucially, we
model $p_i$ as a two-point function of the projected luminosity profile,
\begin{equation}
  p_i = p\!\left(L_{X,500c},\;z,\;T_{\exp},\;\mathrm{EM}(r_p{=}0)\right),
  \label{eq:det_prob}
\end{equation}
which anchors the selection function both at the virial radius (via
$L_{X,500c}$) and at the cluster centre (via the central line-of-sight
emission measure $\mathrm{EM}(r_p{=}0)$, denoted also as ${\rm EM}_0$). This two-point anchor accounts
for the preferential detection of clusters with steeper profiles at fixed
virial luminosity \citep{Ghirardini+22_efeds_morphological_properties}.
The selection function was calibrated by the eROSITA
collaboration\footnote{We thank Nicolas Clerc for providing the eROSITA selection function.} using
image simulations \citep{Liu+21_selection_fct, Comparat+19_efeds_sel_fct,
comparat+20_efeds_sel_fct} that populate dark-matter lightcones with
realistic cluster profiles across the full mass--redshift range of
\efeds, yielding a luminosity scatter of $\sigma_{\log L_X} \approx
0.21$, consistent with the literature \citep{Lovisari+15}. In practice,
$L_{X,500c}$ and $z$ are taken from the \efeds catalogue
\citep{efeds_mass_calibration}, $T_{\exp}$ is derived as
\texttt{ML\_CTS\_0/ML\_RATE\_0} (peaking at $\approx 1200\,\mathrm{ks}$,
consistent with \cite{Bulbul+22_vignetting}), and
$\mathrm{EM}(r_p{=}0)$ is computed from the best-fit $n_e$ profiles of
\cite{bahar_efeds_profiles}.

\paragraph{Step 3: Outlier screening.}
A cluster with an unusually low detection probability $p_i$ receives a
very large weight $w_i$ and can single-handedly dominate its bin. Such
extreme weights may arise from a genuinely rare object detected near the
survey flux limit, but can also result from instrumental artefacts or an
inaccurate virial luminosity estimate. To identify and remove these
outliers, we construct a forward model of the \efeds sample: we populate a
large-area mock survey that reproduces the correct median masses,
redshifts, and virial luminosities in each bin, and record the
distribution of the per-bin maximum weight $\max_i(w_i)$. Any \efeds
cluster whose weight exceeds the 95th percentile of this distribution is
flagged and removed (see App.~\ref{app:forward_model} for details).
After this step, 432 clusters remain
and enter the stacking analysis.

\paragraph{Step 4: SF-weighted median stack.}
The data vector in each bin $k$ is the selection-function-weighted
median luminosity profile,
\begin{equation}
  \hat{L}_X^{(k)}(r_p)
  = \mathrm{med}\!\left[\,L_{X,i}(r_p)\,;\,w_i\,\right]_{i\in k},
  \label{eq:wmed_profile}
\end{equation}
where the weighted median is computed independently at each projected
radius $r_p$. The representative mass and redshift of bin $k$ are the
corresponding weighted medians,
\begin{equation}
  \bar{M}_{500c}^{(k)} = \mathrm{wmed}\!\left[M_{500c,i}\,;\,w_i\right]_{i\in k},
  \qquad
  \bar{z}^{(k)} = \mathrm{wmed}\!\left[z_i\,;\,w_i\right]_{i\in k}.
  \label{eq:wmed_mass_z}
\end{equation}

\paragraph{Step 5: Bootstrap covariance and mass prior.}
The mass uncertainties of individual \efeds clusters are
often comparable to the bin width, so mass errors propagate non-trivially
into the bin assignment and the stacked mass. We propagate these
uncertainties through a single bootstrap over the full catalogue rather
than bin-by-bin. In each of $N_\mathrm{boot}$ realisations $b$:
\begin{enumerate}
  \item[(5a)] We draw a catalogue by SF-weighted resampling with
    replacement (probability $\propto w_i$) and perturb each cluster mass
    by a Gaussian draw from its measurement uncertainty,
    $M_{500c,i} \to M_{500c,i} + \delta M_i$.
  \item[(5b)] Clusters are reassigned to bins and the weighted median
    mass $\MWL^{(b)}$ and cumulative profile $L_{X,\mathrm{cum}}^{(b)}(r_p)$
    are recomputed in each bin.
\end{enumerate}
The variance of the bootstrap mass estimates,
\begin{equation}
  \sigma_{\MWL}^2
  = \frac{1}{N_\mathrm{boot}-1}
    \sum_{b=1}^{N_\mathrm{boot}}
    \!\left(\MWL^{(b)} - \overline{M}_{500c}\right)^2,
  \label{eq:bootstrap_var}
\end{equation}
defines a Gaussian mass-bias prior for each \efeds bin,
\begin{equation}
  b_{\rm \efeds} \sim \mathcal{N}\!\left(0,\;\sigma_{\MWL}^2\right).
  \label{eq:mass_prior}
\end{equation}
The covariance matrix of the full data vector is estimated from the
bootstrap scatter of the profile realisations,
\begin{equation}
  C_{ij}
  = \frac{1}{N_\mathrm{boot}-1}
    \sum_{b=1}^{N_\mathrm{boot}}
    \!\left(L_{X,i}^{(b)} - \bar{L}_{X,i}\right)
    \!\left(L_{X,j}^{(b)} - \bar{L}_{X,j}\right),
  \label{eq:bootstrap_cov}
\end{equation}
where indices $i,j$ run over all radial bins in all mass--redshift bins.
Cross-bin correlations (between different mass--redshift bins) are found
to be negligible and are set to zero in the remainder of this work.

The purity of all our bins is above 80\%, estimated as a fraction of objects in a given bin with \texttt{FCONT}$<0.3$ \cite{efeds_optical_confirmation}. This value is compatible with the simulation-based estimates of \cite{comparat+20_efeds_sel_fct}.

\section{Inference}
\label{sec:inference}

We aim to use the measured stacked surface brightness profiles of the CHEX-MATE clusters and the rest-frame soft-band luminosity profiles of the \efeds sample to constrain baryonic feedback. In Sec.~\ref{sec:modelling}, we introduced the parametric model based on the baryonification framework, which can connect three-dimensional density profiles of halos with the \CR (\lx) profiles, the gas fraction as a function of halo mass, or even the suppression of the matter power spectrum due to baryons. In this work, we proceed in two steps, both based on our BFC modelling: first, we measure gas fractions by fitting the surface brightness (or luminosity) profiles of the \CM and \efeds data in individual bins; second, we jointly fit the resulting gas fractions across multiple bins (i.e., at different masses and redshifts). In upcoming work, we will extend this approach to jointly fitting the X-ray profiles directly.

We perform parameter estimation within a Bayesian framework, exploring the posterior distribution of model parameters using Markov Chain Monte Carlo (MCMC) sampling. We employ a Gaussian likelihood function, $\mathscr{L}$, defined as
\begin{equation}
    \label{eq:likelihood}
    \log \mathscr{L} = -\frac{1}{2} 
    \left(\mathbf{d}_{\rm BFC} - \mathbf{d}_{\rm obs} \right)^T
    C^{-1}
    \left(\mathbf{d}_{\rm BFC} - \mathbf{d}_{\rm obs} \right) + \mathrm{const}.,
\end{equation}
where the data vector $\mathbf{d}_{\rm BFC}$ corresponds to the model predictions, $\mathbf{d}_{\rm obs}$ denotes the data, and $C$ is the covariance matrix.
For the fits to the binned X-ray profiles, we model $\SB \equiv \CR(\theta)$ computed using Eq.~\eqref{eq:CR_Int} (for \CM) and $L_{X,\rm cum}$ using Eq.~\eqref{eq:lx_cum} (for \efeds). When fitting to gas fractions instead of the X-ray profiles, the data vector $\mathbf{d}_{\rm obs}$ consists of gas-fraction values at different masses.
When fitting to gas fractions, we convert the BFC profiles into gas-fraction values to define $\mathbf{d}_{\rm BFC}$, and compare these to the measured gas fractions reconstructed from the per-bin fits. In the latter case, we assume that data from different mass and redshift bins are independent, with uncertainties recovered from the preceding MCMC analysis.

To infer the BFC model parameters, we run an MCMC using the \texttt{emcee} package\footnote{\url{https://emcee.readthedocs.io/}}. Specifically, we employ 20 walkers that explore the parameter space in parallel. The dimensionality of the parameter space depends on the observational data being fitted. In the case of a single stacked profile, i.e., data at a single mass and redshift, we replace $M_c$ and $\mu$ (which parameterise the feedback strength as a function of mass) with a single slope parameter $\beta$, sampled directly instead of assuming the functional form $\beta(M_{200c}, M_c, \mu)$.
When fitting to data across multiple mass bins, we typically employ the full functional form $\beta(M_{200c}, M_c, \mu)$ with two independent parameters $M_c$ and $\mu$ in the MCMC chain. We provide an overview of the BFC parameters and their assumed prior ranges in Tab.~\ref{tab:model_params}. The priors are selected to be wide enough to encompass most physically plausible feedback scenarios. When deriving the gas fractions for individual \efeds bins, we increase the priors on $\beta$ and $\delta$ to $(0,6)$, and $(-3, 11)$, respectively, making sure the resulting measurements are not prior-limited. 

\begin{figure}
    \centering
    \includegraphics[width=1.\linewidth]{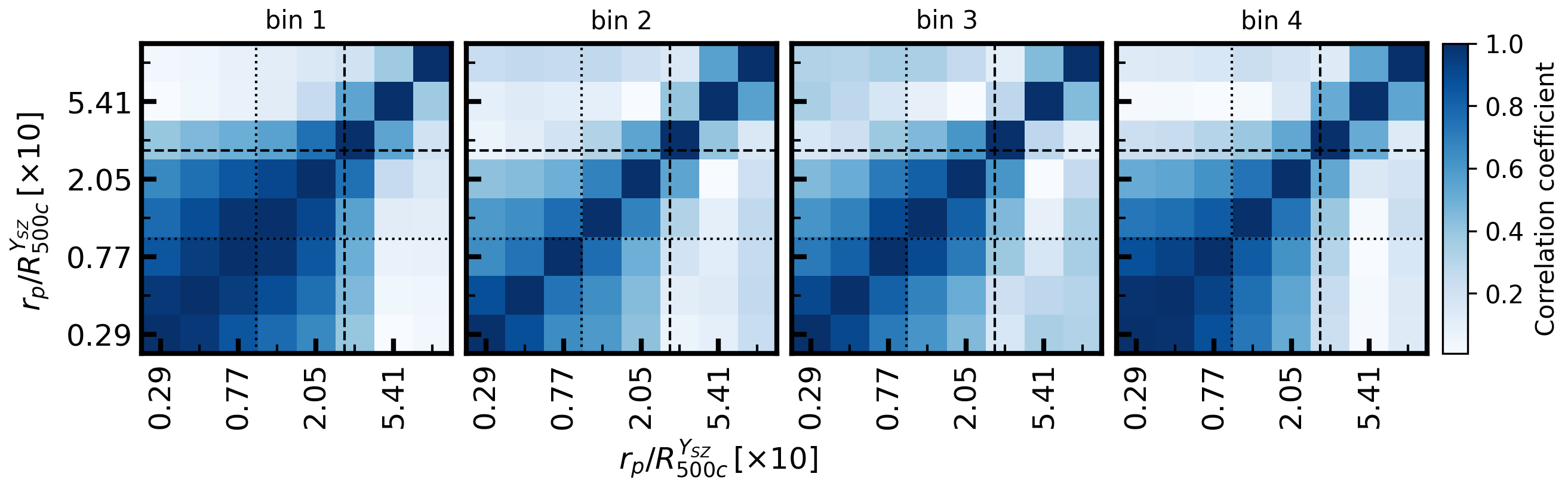}
    \caption{Correlation matrices for CHEX-MATE \CR profiles in four different bins (see Tab.~\ref{tab:chm_bins}), obtained via the bootstrap method. The dotted and dashed lines correspond to $r_p=0.1\RSZ$ and $r_p=0.3\RSZ$, respectively.}
    \label{fig:covariances_bins1234}
\end{figure}

\begin{table}[h!]
    \renewcommand{\arraystretch}{1.2} 
    \centering
        \begin{tabular}{l p{10cm} l}
        \hline
        \hline
        \textbf{Parameter} & \textbf{Description} & \textbf{Prior Range} \\
        \hline
        $\log_{10} M_c$    & Characteristic mass scale of feedback ($\beta = 3/2$) & $\mathcal{U}[11,15]$ \\
        $\mu$              & Sharpness of the feedback mass scaling around $M_c$ & $\mathcal{U}[0,2]$ \\
        $\beta$            & Intermediate slope of the gas density profile; used for single-bin inference instead of $(M_c, \mu)$ & $\mathcal{U}[0,3]$ \\
        $\theta_{\rm co}$  & Core radius size in units of $r_{200c}$ & $\mathcal{U}[0.001,0.5]$ \\
        $\delta$           & Outer slope of the gas density profile & $\mathcal{U}[1,11]$ \\
        $b_{\rm SZ}$       & SZ mass bias parameter & $\mathcal{U}[-0.5,-0.1]$ \\
        
        $dT_{\rm \efeds}$       & Temperature uncertainty of \efeds objects & $\mathcal{N}[0.0,0.4]$ \\
        
        $Z_{\rm \efeds}$       & Metallicity of \efeds objects & $\mathcal{N}[0.3Z_\odot,0.15Z_\odot]$ \\
        
        $b_{\rm \efeds}$       & Mass uncertainty of \efeds bins & $\mathcal{N}[0.0,\delta M_{500c}]$ \\
        \hline
        \end{tabular}
    \caption{Free parameters of the BFC X-ray model, their descriptions, and prior ranges.}
    \label{tab:model_params}
\end{table}

\section{Model validation and consistency}
\label{sec:validation}

\begin{figure}[h]
    \centering
    \includegraphics[width = \textwidth]{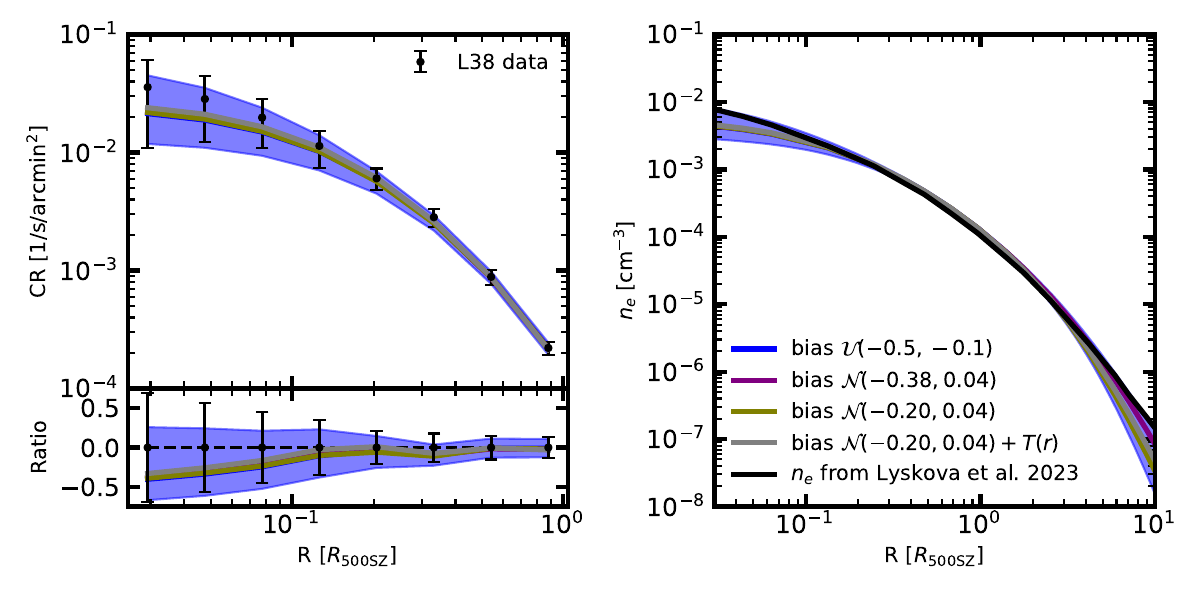}
    \caption{Fit to the surface brightness of the 38-cluster subsample of CHEX-MATE defined in \cite{Lyskova_2023}, referred to as \ls. In the left panel, we show the 95\% confidence interval from the MCMC chain fit to the data, shown in black. In the right panel, we translate the posterior distribution into electron number density profiles. We compare the resulting profile to the one presented in \cite{Lyskova_2023}.}
    \label{fig:validation_ne_lyskova}
\end{figure}

\begin{figure}[h]
    \centering
    \includegraphics[width = \textwidth]{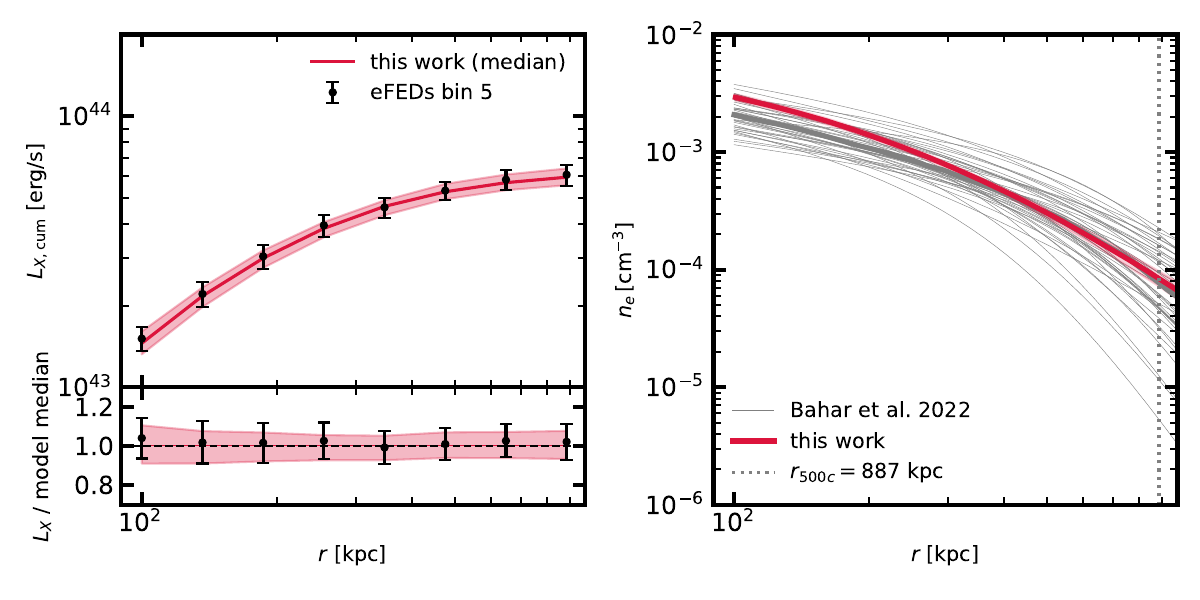}
    \caption{Fit to the luminosity profile of \efeds bin 5. The left panel shows the fitted data vector (black), while the red line and band show the MCMC chain median and 16-84 percentiles, respectively. The bottom-left panel shows the ratios relative to the MCMC median. In the right panel, we show the predictions of our analysis for the three-dimensional electron density profiles (median + 16-84 percentiles shown in red) and compare them to the electron profiles of the individual clusters of the same bin obtained by \cite{bahar_efeds_profiles} (thin grey lines) and their median (thick grey line).}
    \label{fig:validation_ne_efeds}
\end{figure}

\noindent Before inferring constraints on baryonic feedback from the \CM and \efeds samples, binned as a function of mass and redshift, it is important to test the internal consistency of the data considered in this analysis. We begin by validating our modelling of the stacked X-ray profiles of \CM clusters. For this purpose, we use the \ls subsample of the full \CM sample presented in \cite{Lyskova_2023} and introduced in Sec.~\ref{sec:data}. This choice is motivated by the availability of the three-dimensional electron number density profile for the \ls sample. This profile was obtained by fitting a projected Vikhlinin profile to the stacked surface brightness profiles of \ls objects extracted from SRG/eROSITA images in the 0.3--2.3~keV band. To obtain the median CR profile for the sample, we match all clusters presented in \cite{Lyskova_2023} within the full \CM catalogue and compute the median stacked profile, while the electron density profile is taken directly from \cite{Lyskova_2023} (see their Fig.~4).

In the left panel of Fig.~\ref{fig:validation_ne_lyskova}, we show the median CR profile of the \ls sample as black data points, where the error bars represent the diagonal elements of the covariance matrix constructed via the bootstrap method (see Sec.~\ref{sec:inference} for details). In the top-left panel, we show the best-fit models for different hydrostatic mass-bias modelling choices. The solid blue line represents the best fit assuming a flat prior on the hydrostatic mass bias, $\mathcal{U}[-0.5,-0.1]$, which encompasses most literature results, with the envelope depicting the 95\% confidence interval obtained from the MCMC sampling. The solid purple and olive lines are obtained from the best-fit \CR{} assuming a Gaussian prior on the mass bias found by \cite{Sereno_2025}, and a Gaussian prior with the same standard deviation as found by \cite{Sereno_2025} but a mean of $-0.2$, respectively. The latter agrees with the lower mass-bias values found by several weak-lensing studies \cite[][and references therein]{wl_masses}. The grey line was obtained by assuming the same Gaussian prior on the mass bias as in the previous scenario, but replacing the average temperature $T_{\rm avg}$ of the cluster bin with the radial temperature profile $T(r)$ as provided by the BFC framework. In the bottom left panel, we show the ratios between the best-fit model predictions and the CHEX-MATE data for the \ls sample. Taking the best-fit BFC parameters, we compute the three-dimensional electron density profiles $n_e$ for the different mass bias scenarios, shown as solid coloured lines in the right panel of Fig.~\ref{fig:validation_ne_lyskova}, and compare them to the profile constructed in \cite{Lyskova_2023} from the same \ls sample, plotted as a solid black line. As can be seen, the BFC model — derived by fitting the surface brightness profiles — provides a good description of the three-dimensional electron number density, a quantity it was not fitted to, when compared with the literature. This illustrates the self-consistency of the model developed in this work and provides a more direct test than, e.g., comparing gas fractions, since surface brightness measurements indirectly probe the projected gas density and not only the total mass. Additionally, as seen in Fig.~\ref{fig:validation_ne_lyskova}, the results are only slightly affected by the specific choice of mass bias, which would not be the case for gas fractions, where the total mass inside the virial radius plays an important role (see App.~\ref{app:bias_effects}). Note that the density profile from \cite{Lyskova_2023} is also model-dependent, as the reconstruction of the three-dimensional profile from the stacked X-ray data assumes a specific parametric form for the electron profile. Our result confirms that these two approaches yield consistent results. In addition, we include more details on the temperature modelling in Appendix~\ref{app:T_modelling}. In addition, we compared the inferred gas mass inside the virial radius, obtained from a fit to the stacked profile of bin~4, to the preliminary result of the \CM collaboration, and found agreement better than 5\%.

We perform a similar exercise for the \efeds sample. We assume bin~5 (low-z, high-mass), as this bin is least biased by the X-ray selection, and perform a fit to the cumulative luminosity profile. We show the \efeds data vector in the left panel of Fig.~\ref{fig:validation_ne_efeds} as black data points and the fit in red; the corresponding residuals are shown in the bottom panel. In the right panel, we compare the electron number density profile reconstructed from our MCMC analysis (red) with the Vikhlinin-reconstructed profiles of objects in bin~5 from \cite{bahar_efeds_profiles}. Overall, while the two approaches are based on different models, we find good agreement between them. 

Last but not least, for lower-mass, group-scale objects, the X-ray emissivity depends on temperature much more strongly than it does at cluster scales. We make sure that our choices regarding temperature and metallicity modelling are conservative enough; see App.~\ref{app:T_modelling} for more details.


\section{Results}
\label{sec:results}

\begin{figure}
    \centering
    \includegraphics[width=1\linewidth]{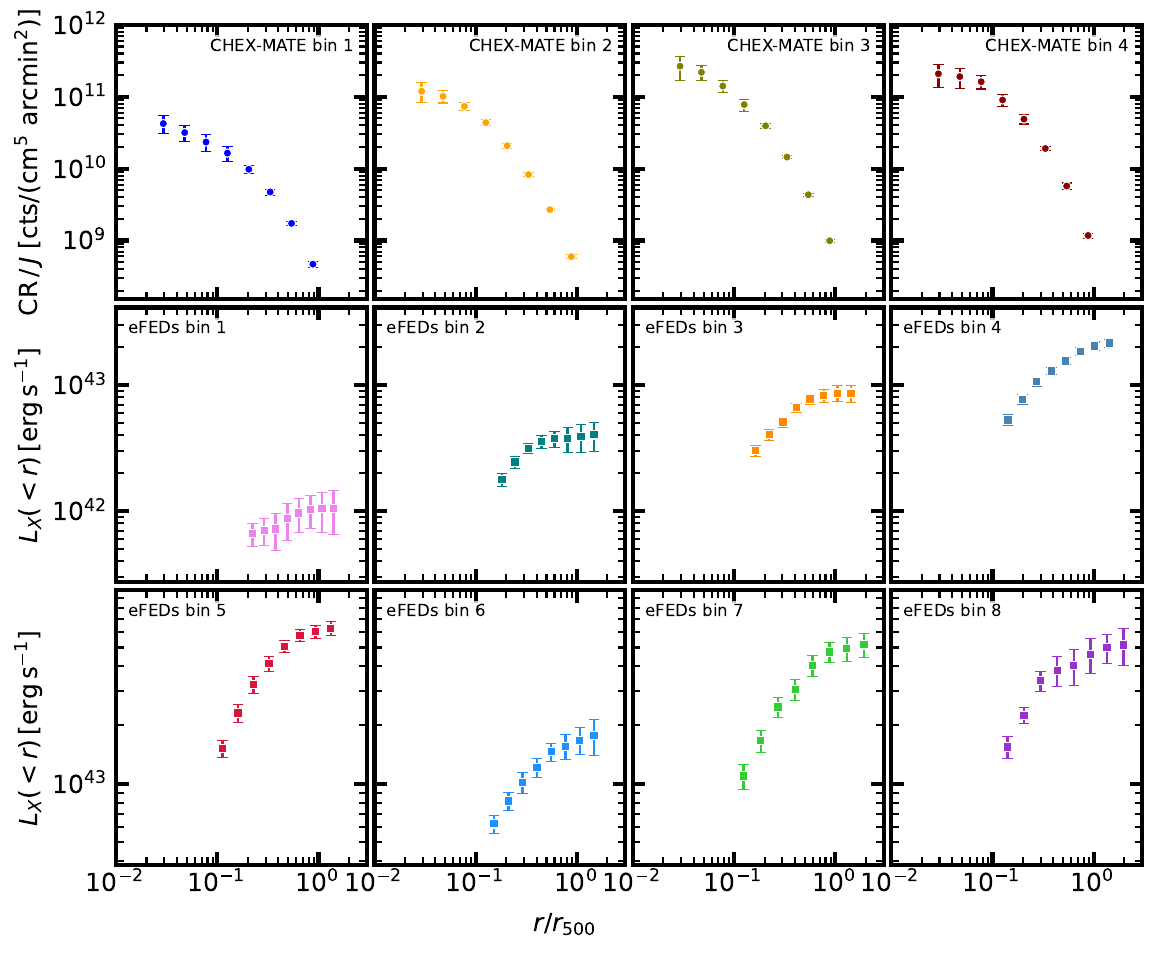}
    \caption{Data vector: the panels in the first row show \CM bins~1-4, specifically the differential surface brightness profiles \CR \eqref{eq:CR_Int} divided by X-ray emissivity $J$ \eqref{eq:J}, thus 2-dimensional emission measure profiles as a function of radius rescaled by \RSZ. The central and the bottom rows display the cumulative luminosity profiles of 8 \efeds bins. The profiles are plotted as a function of $r/r_{500c}$.}
    \label{fig:data_vectors}
\end{figure}

In Fig.~\ref{fig:data_vectors} we show the outcome of our binning procedure for both the \CM and \efeds samples. The top row shows the $\CR/J$ profiles (surface brightness profiles divided by the X-ray emissivity) of the \CM sample, while the middle and bottom rows show the cumulative luminosity profiles of the \efeds sample (colours are consistent with the binning overview in Fig.~\ref{fig:chm_bins}. These are the final data products used to measure the gas fractions in our analysis. 

\begin{figure}
    \centering
    \includegraphics[width=0.8\linewidth]{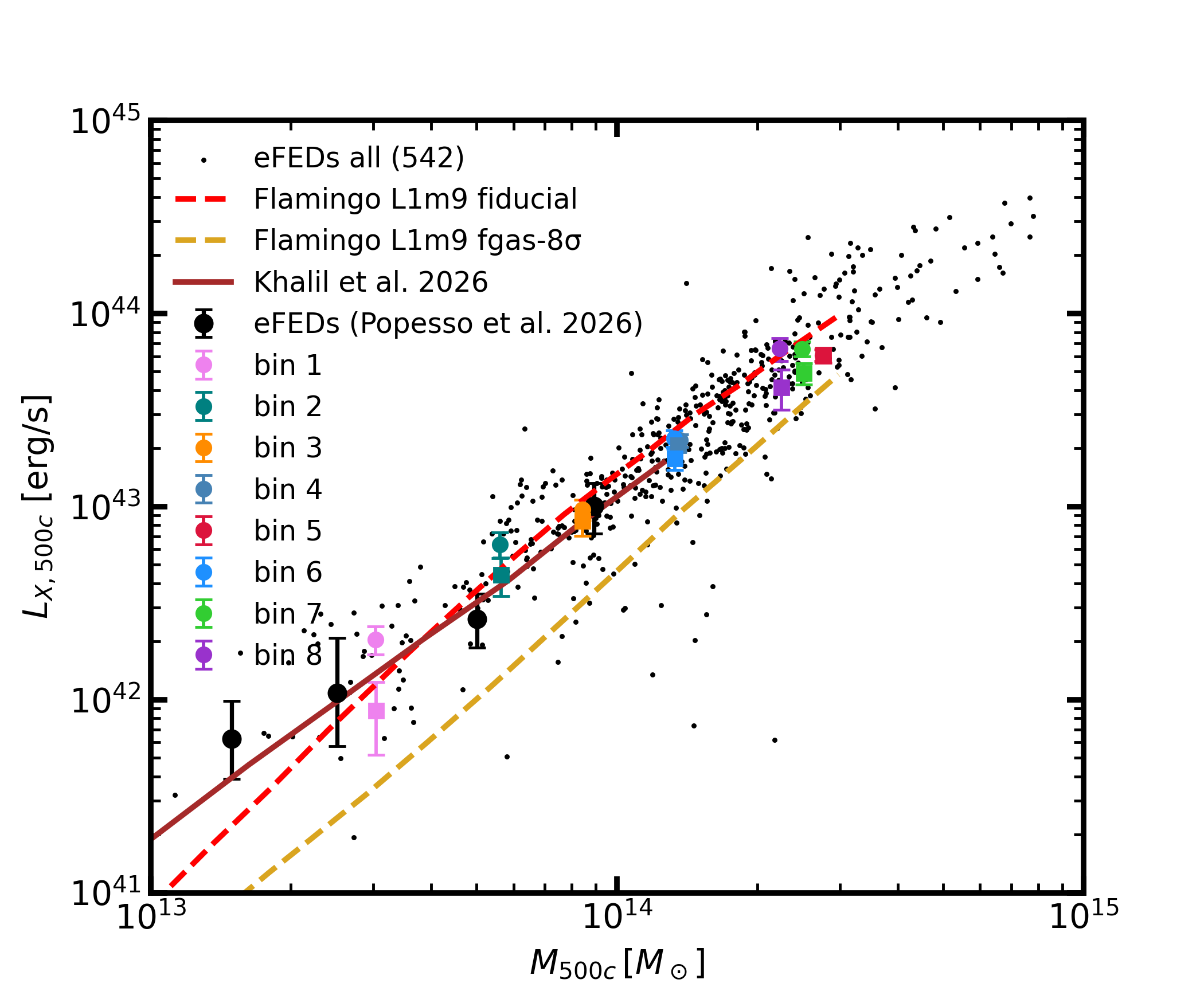}
    \caption{\efeds $L_X-M$ scaling relation. We show masses and luminosities of all \efeds objects at $r_{500c}$ as measured by \cite{bulbul_24_erosita} as black points and the binned optically selected \efeds objects of \cite{Popesso_2024} as black data points with error bars. The coloured data points represent values measured from our binned profiles. Circles denote binning without accounting for selection function effects, while the squares show the $L_x-M$ values of our 8 bins after including the effects of the selection function. The solid brown line shows the mass-luminosity relation found by \cite{Khalil+26_erosita}.}
    \label{fig:M-Lx-scaling-relation}
\end{figure}

In Fig.~\ref{fig:M-Lx-scaling-relation} we show the mass-luminosity relation of the \efeds objects at virial radius. The small black circles represent the $M_{500c}$ and $L_{X,500c}$ of the individual \efeds objects. The coloured points show the median masses and luminosities of the binned sample together with the uncertainty given by the 16\%-84\% range of the values in a specific bin. The circles represent values with no selection-function effects accounted for, while squares represent values after weighting the objects with the inverse detection probabilities given by the selection function. We observe that bins closer to the detection limit (for example, bins~1 and 2) are severely affected by selection, while bins~3 and 4, well above the instrument's detection limit, are essentially unbiased. The values are broadly consistent with the results of \cite{Popesso_2024} and, especially at lower masses, agree better with the scaling relation predicted by the fiducial feedback scenario of Flamingo suite than with the 'strong' feedback scenario, similar to the findings of \cite{eckert_2025arXiv251204203E} and \cite{Khalil+26_erosita}.

\begin{figure}
    \centering
    \includegraphics[width=1\linewidth]{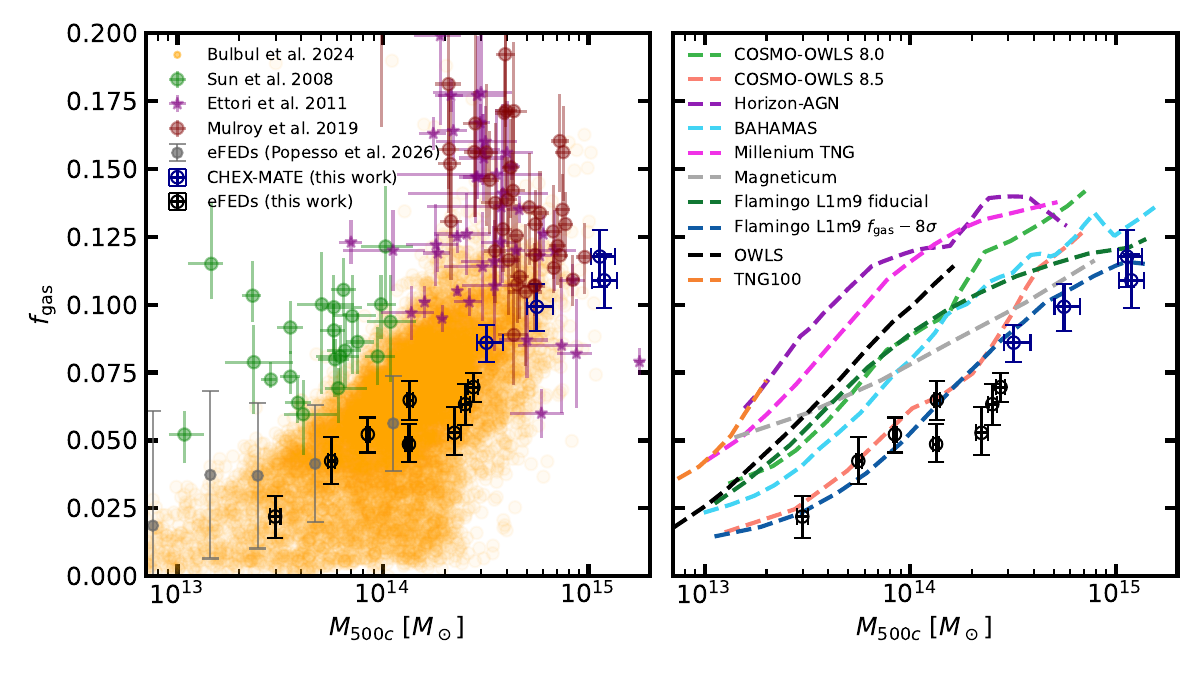}
    \caption{Gas fractions of the binned \efeds (blue) and \CM (black) catalogue as a function of total mass as found in this work compared to a compilation of past gas fraction measurements (left panel) and to the gas fractions recovered from a range of hydrodynamical simulations (right panel).}
    \label{fig:fgas_data}
\end{figure}

\begin{figure}
    \centering
    \includegraphics[width=1\linewidth]{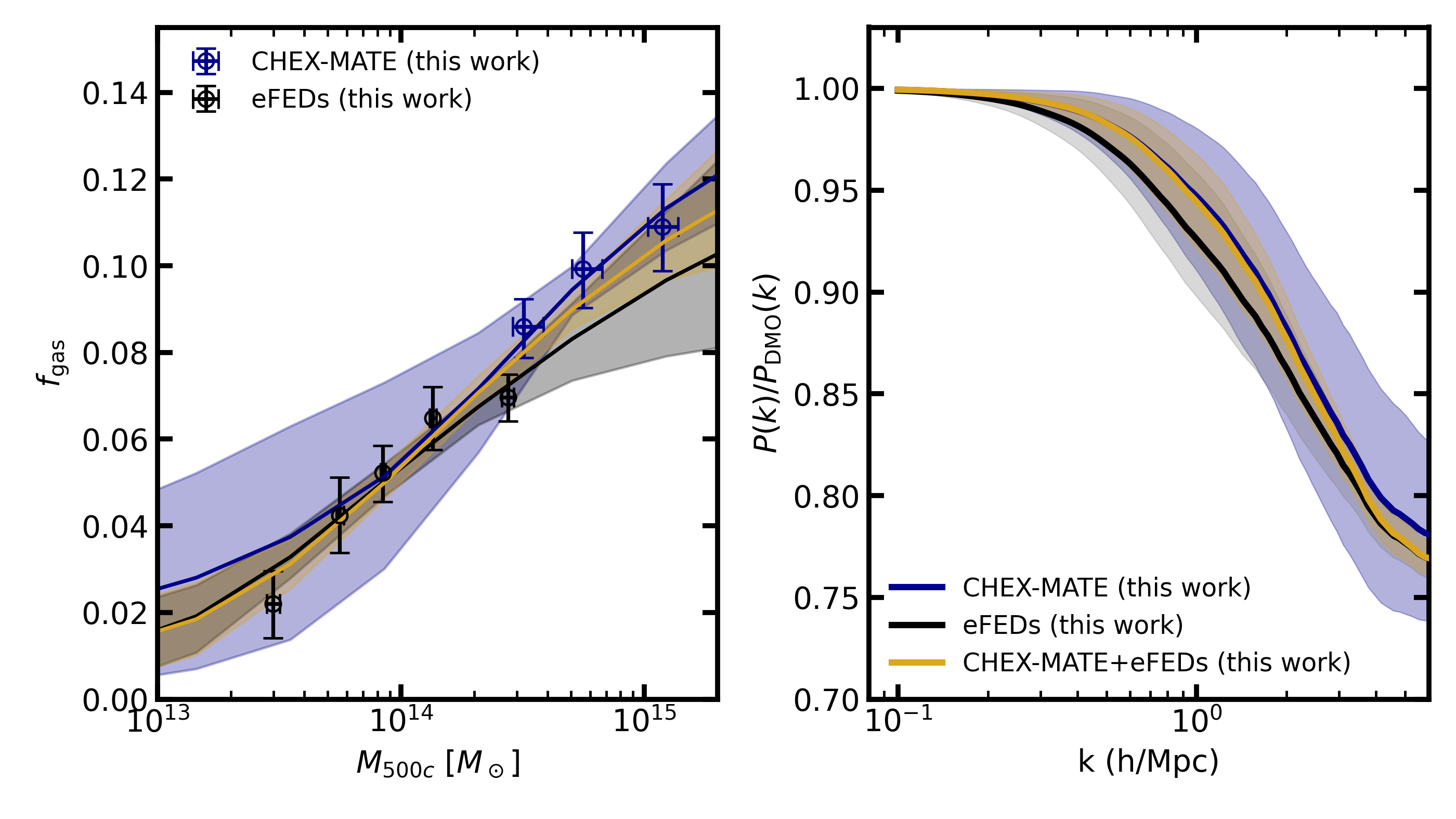}
    \caption{Gas fractions and power spectrum suppression from the fits to \efeds and \CM low-redshift gas fraction data points measured in this work. In the left panel, the dark blue line and envelope represent the median and 16\%-84\% confidence intervals of the BFC fit to the low-z \CM data points (dark blue), while the black line and grey envelope show the same for fits to the \efeds data shown as dark points with error bars. The golden colour shows the result of a combined \efeds + \CM fit. In the right panel, we show the matter power spectrum suppression for the three scenarios as recovered by the emulator introduced in Sec.~\ref{subsec:modelling_pk_emulator}.}
    \label{fig:fgas_from_fgas}
\end{figure}

\begin{figure}
    \centering
    \includegraphics[width=1\linewidth]{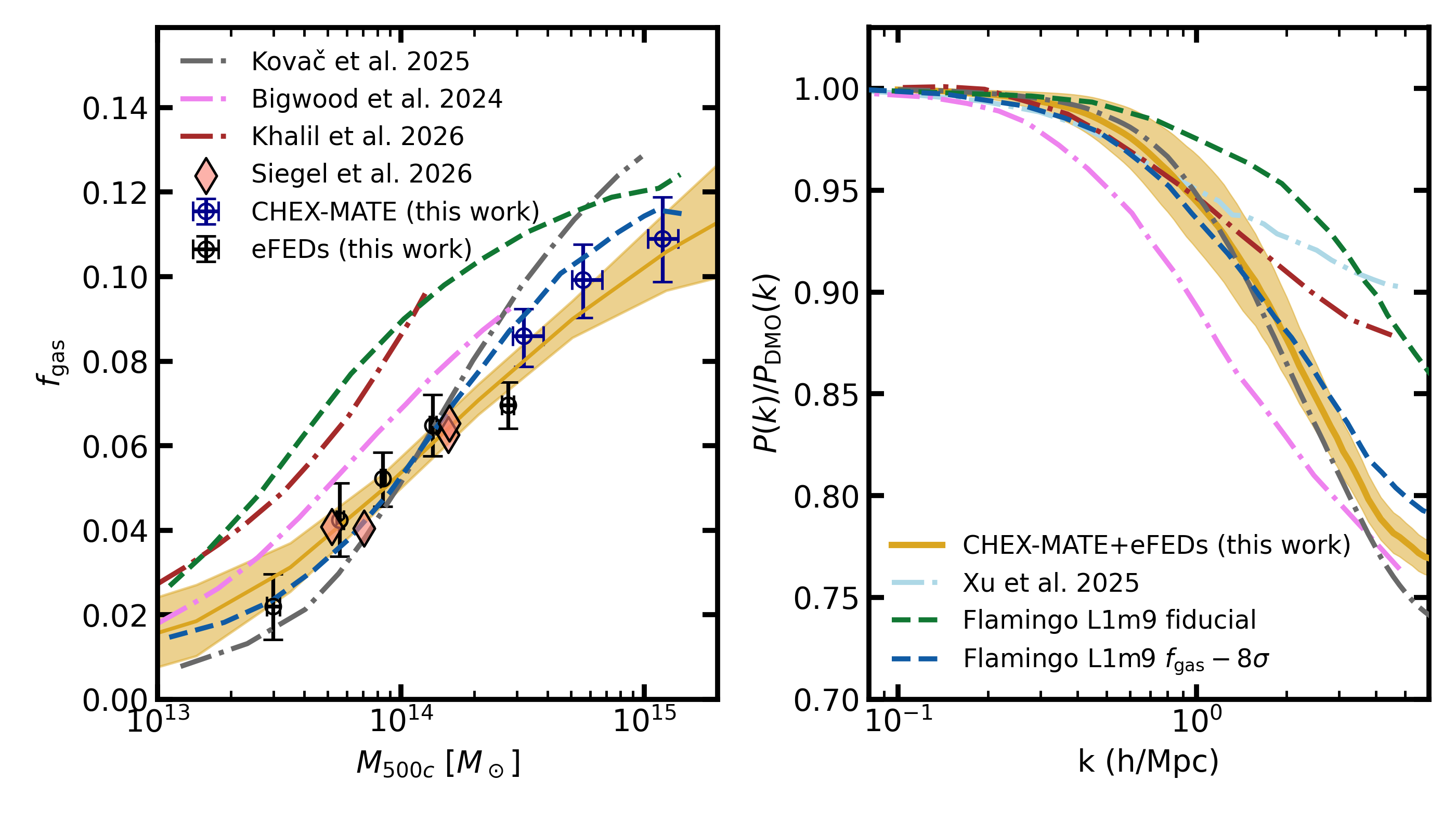}
    \caption{Comparison of our gas fraction measurement (black and dark blue points in the left panel) and the joint fit to these data (golden in the left panel) to a collection of recent results from the literature. In the right panel, we show the matter power spectrum suppression from the joint fit to the gas fractions, with the golden line showing the median suppression and the envelope showcasing the 16\%-84\% confidence interval. The dashed green and blue lines show the corresponding quantities from the Flamingo simulations implementing fiducial and strong feedback, respectively. For comparison, we add the recent gas fractions and power spectrum constraints from \cite{Kovač_2025} (grey), \cite{Bigwood_2024} (violet), \cite{Khalil+26_erosita} (brown), \cite{xu_arxiv} (light blue) and \cite{siegel_2025} (red diamonds).}
    \label{fig:fgas_from_fgas_compare}
\end{figure}

In Fig.~\ref{fig:fgas_data}, we present the gas fraction as a function of halo mass, as inferred from our fits to the X-ray profiles of \efeds (black) and \CM (dark blue). The data points and uncertainties are obtained by sampling the posteriors from our fits. For each sampled MCMC configuration, we compute the total mass and the gas fraction values using Eq.~\eqref{eq:fgas}, where the gas and total mass profiles are obtained from the BFC model. We display the median values with 16\%-84\% quantiles from the sampled MCMC points. In the left panel of the same figure, our results are compared to a compilation of gas fraction measurements\footnote{We thank Giovanni Aricò for providing these.}. In grey, we show the results from optically selected \efeds objects of \cite{Popesso_2024}, and in yellow, the individual eRASS:1 gas fractions \cite{bulbul_24_erosita}. In addition, we show gas fractions from 43 groups observed by Chandra \cite{Sun+09_fgas_chandra_groups}, measurements obtained using 44 luminous clusters from XMM-Newton \cite{Ettori+11_fgas_XMMNewton} and from 41 X-ray-luminous clusters observed with Chandra/XMM \cite{mulroy+19_fgas_chandra_xmm}. In the right panel, we compare our measurements to gas fractions predicted by several hydrodynamical simulations, specifically cosmo-OWLS \cite{lebrun+14_cosmoowls}, Horizon-AGN \cite{Dubois+14_horizon_agn_main_paper}, BAHAMAS \cite{mccarthy+17_bahamas}, Millenium-TNG \cite{pakmor+23_mtng}, Magneticum \cite{hirschmann+14_magneticum}, Flamingo \cite{flamingo}, OWLS \cite{Schaye+10_OWLS} and TNG100 \cite{Phillepich+18_TNG}. In the overlapping mass range (\MWL = $3\times 10^{13} - 10^{14}~M_\odot$), our measurements agree with the stacks of optically selected \efeds objects \cite{Popesso_2024} and with the gas fraction measurements of the X-ray-selected eRASS:1 sample \cite{bulbul_24_erosita}. On the other hand, the Chandra groups \cite{Sun+09_fgas_chandra_groups} and the luminous clusters of \cite{Ettori+11_fgas_XMMNewton} and \cite{mulroy+19_fgas_chandra_xmm} give considerably larger gas fractions, potentially due to selection effects. Especially, the first two of the cited studies do not correct for the effects of selection; therefore, the resulting systems have, on average, larger gas fractions than the overall population. Compared to hydrodynamical simulations, our high-mass measurements of \CM align well with the results from the strong-feedback FLAMINGO simulation ($f_{\rm gas}-8\sigma$ variant). The four low-mass, low-redshift ($z<0.4$) bins also show very good agreement with the strong-feedback FLAMINGO run, while the most massive low-redshift bin, as well as three high-redshift bins ($z>0.4$), hint at even stronger feedback. 

We now proceed to fit the inferred gas fractions from Fig.~\ref{fig:fgas_data} using a single global BFC model. In this part, we restrict ourselves only to the low-redshift measurements to focus primarily on the mass dependence of feedback. Specifically, we assume $f_{\rm gas}$ values from bins~1-5 of \efeds ($z<0.4$) shown as black data points and bins~1-3 of the \CM sample ($z<0.33$) plotted as dark blue data points in the left panel of Fig.~\ref{fig:fgas_from_fgas}, in which we investigate the joint fits to the measured gas fractions from both samples. When fitting the data, we replace the $\beta$ parameter with $M_c$ and $\mu$, thus injecting the mass dependence into the BFC feedback parametrisation. The dark blue (black) shaded area illustrates the 68\% confidence interval (C.I.) from the MCMC fit to the \CM (\efeds) low-redshift data points. Overall, the fit matches the measurements well, with reduced constraining power at lower (higher) masses, where the fit extrapolates. The BFC parameters recovered from both fits are mutually compatible (see Fig.~\ref{fig:fgas_contours} of App.~\ref{app:mcmc_posteriors}), thus we combine all 8 gas fraction points into a single measurement and fit them jointly. The result is shown as a golden line (median) and envelope (68\% C.I.). In the right panel of the same figure, we translate the gas-fraction fits into constraints on matter power spectrum suppression, with \CM giving weaker constraints than the \efeds and \efeds+\CM scenarios. This is connected to the mass range probed by both samples. The \efeds sample probes better the halo masses of $10^{13}-10^{14}M_\odot$, relevant for matter power spectrum scales affected by baryons, while \CM sample does not probe low enough halo masses to constrain feedback above $k\sim 1h/\rm Mpc$ significantly.

In Fig.~\ref{fig:fgas_from_fgas_compare}, we compare the results of our joint fit analysis to a selection of results from hydrodynamical simulations and studies based on the actual observations. Consistent with the gas-fraction result, our power-spectrum suppression generally agrees well with the strong-feedback Flamingo simulation (shown in dashed blue). The grey line shows the constraints from the ACT DR5 kinematic (kSZ) SZ measurement stacked around BOSS CMASS galaxies combined with the gas fractions from \efeds sample of \cite{Popesso_2024}, as published by \cite[][with multiple authors of this work contributing]{Kovač_2025}. The authors find slightly stronger feedback at low masses, while hinting at less efficient feedback in the largest clusters compared to our result. The DES Y3 + kSZ constraints of \cite{Bigwood_2024} (violet, which also varies cosmological parameters) show slightly larger gas fractions, while yielding a stronger suppression of the matter power for $k>0.1\,h/\rm Mpc$. As light red diamonds, we show four $f_{\rm gas}$ values measured from the eRASS:1 sample by \cite{siegel_2025}, which are in agreement with our results. On the other hand, \cite{xu_arxiv} (light blue) measures a more moderate feedback using DES Y3 and Planck PR4 data. As a last example, we show a recent result from \cite{Khalil+26_erosita}, who analysed 25 low-redshift groups in the eRASS:1 catalogue (brown line in both panels). The authors find less extreme feedback, aligning better with the fiducial feedback prescription of the Flamingo suite (dashed red curves) at the level of gas fractions. Their suppression of the power spectrum, however, is more compatible with the strong feedback FLAMINGO implementation below $k\lesssim 2\,h/\rm Mpc$, while approaching fiducial Flamingo towards $k\sim 5\,h/\rm Mpc$.

\begin{figure}
    \centering
    \includegraphics[width=1.\linewidth]{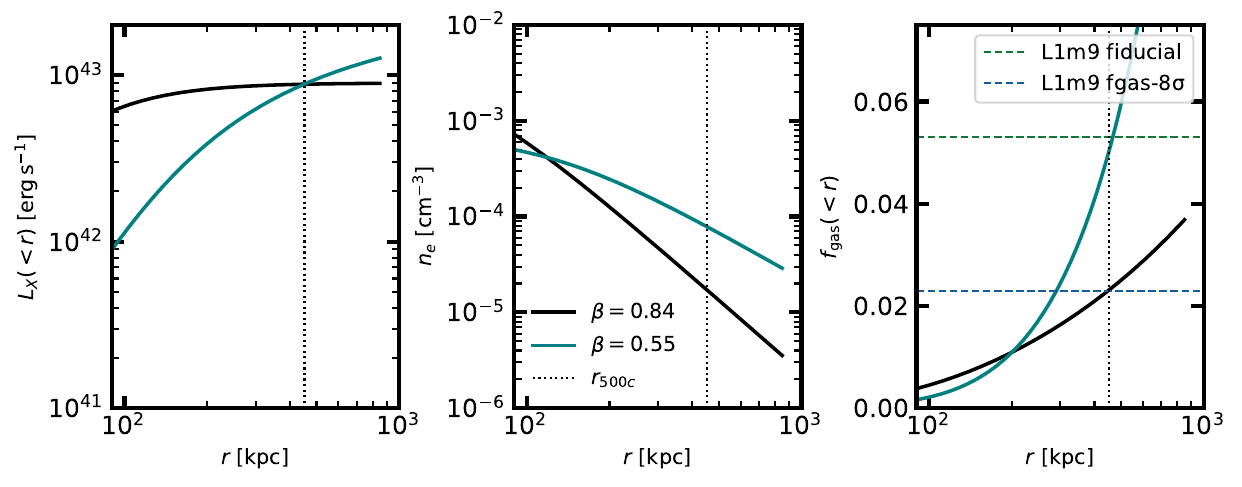}
    \caption{The effects of the X-ray profile shapes on the gas fractions at a fixed point along the mass-luminosity relation. The left panel shows two examples of cumulative luminosity profiles with the same $L_{X,500c}=8.8\times 10^{42}$~erg/s and $M_{500c} = 3.04\times 10^{13}\,M_\odot$, although having different shapes. The middle panel shows the corresponding three-dimensional electron density profiles and the right panel displays the cumulative gas fractions, all as a function of radius. The vertical dotted line highlights the virial radius of the object and green (blue) horizontal lines mark the gas fraction values at the assumed halo mass under the fiducial (strong) feedback scenario.}
    \label{fig:shapes_effects_lx_fgas}
\end{figure}

Before concluding, we highlight an interesting aspect apparent from Fig.~\ref{fig:M-Lx-scaling-relation} and Fig.~\ref{fig:fgas_from_fgas}. Namely, the fact that it is difficult to make a statement about the strength of baryonic feedback from a measured mass-luminosity scaling relation alone. While our result, the previous \efeds results of \cite{Popesso_2024} and a recent result of \cite{Khalil+26_erosita} all arrive at a similar $M - L_X$ scaling relation, more compatible with moderate (fiducial) Flamingo feedback, the gas fractions and $P(k)$ suppression in Fig.~\ref{fig:fgas_from_fgas} are not similar anymore. While ours and the result of \cite{Popesso_2024} align better with the strong feedback models, \cite{Khalil+26_erosita} remains more compatible with the more moderate feedback variants (at least at the level of the gas fractions). We show in Fig.~\ref{fig:shapes_effects_lx_fgas} that one can obtain agreement with both variants starting from the same mass-luminosity scaling relation if the shapes of the X-ray profiles are different. Since the $L_X$ profile scales as $n_e^2$, while the gas fraction scales as $n_e$, profiles with different shapes do not affect the two quantities equally. We illustrate this with a simple toy model in which we consider two cumulative luminosity profiles sharing the total mass ($M_{500c} = 3\times 10^{13}\,M_\odot$) and virial luminosity $L_{X,500c}=8.8\times 10^{42}\,\rm erg/s$. We assume a two beta-profiles parametrised as $n_e = n_0 \left[1 + (r/r_c)^2\right]^{-3\beta/2}$, a steeper one with $\beta  = 0.84, \,r_c = 50.0 \,{\rm kpc},\, n_0  =   4.45\times 10^{-3}$ (black) and a flatter profile with $\beta  = 0.55, \,r_c = 132.3 \,{\rm kpc},\, n_0  =   7.10\times 10^{-4}$ (teal). The left panel of Fig.~\ref{fig:fgas_from_fgas} shows the two cumulative luminosity profiles, intercepting at $r_{500c}$ (dotted black line). The middle panel shows the corresponding $n_e$ profiles and finally, in the right panel we plot the cumulative gas fractions for the two profiles. We see that the steep profile (black) recovers the gas fraction value of the strong feedback Flamingo simulation ($f_{\rm gas} = 0.023$, dashed blue line), while the flatter (teal) profile aligns with the fiducial feedback gas fraction ($f_{\rm gas} = 0.053$, dashed green line). Indeed the $n_e$ profile of \cite{Khalil+26_erosita} drops as $\sim r^{-1.2}$ at $M_{500c}=2.5\times 10^{13}\,M_\odot$, while our profile at similar mass decreases as $\sim r^{-2}$ between $r  = [0.15-1.0]\,r_{500c}$. As shown above, this might explain a similar mass-luminosity relation while having different gas fractions (and $P(k)$ suppressions) between this work and \cite{Khalil+26_erosita}.

\section{Conclusion}
\label{sec:conclusion}

In this work, we developed a flexible and physically motivated model for stacked X-ray observables of galaxy clusters and groups, built on the baryonification (BFC) framework. By linking the three-dimensional gas density and temperature profiles predicted by BFC to the instrument-convolved X-ray emissivity, our model connects surface brightness and luminosity profiles directly to gas fractions and to the baryonic suppression of the matter power spectrum within a single, self-consistent halo-based description. This provides a modelling framework that extends much closer to the raw X-ray observables than gas-fraction-based approaches, and one that is naturally suited to a joint interpretation of X-ray data alongside other probes of the intracluster medium, such as the Sunyaev-Zeldovich effect.

We first validated the model and tested its internal consistency using two independent subsamples with available three-dimensional density information: the \ls subsample of \CM, and bin~5 of \efeds. In both cases, fitting only the two-dimensional (surface brightness or luminosity) profiles yielded three-dimensional electron density profiles in good agreement with those reconstructed independently in the literature, a quantity the model was not directly fitted to. For \ls, this agreement is essentially unaffected by the choice of hydrostatic mass-bias prior or by whether a constant or radially resolved temperature is assumed. This cross-check is more stringent than a comparison based on gas fractions alone, since it tests the full radial shape of the recovered density profile rather than just its integral, and it demonstrates that our approach is robust to the main modelling choices entering the temperature, metallicity and mass calibration.

Building on this validated model, we measured, for the first time, gas fractions of the \CM sample in four mass-redshift bins and performed similar measurements for the \efeds sample in eight bins spanning the group-to-cluster mass range, explicitly accounting for the effects of X-ray selection. In the overlapping mass range, our measurements agree well with previous gas fraction estimates from optically selected \efeds clusters and from the eRASS:1 sample, while lying below gas fractions reported for X-ray-selected literature samples, some of which are affected by Malmquist-type bias. Compared to hydrodynamical simulations, our results across both samples are consistently best matched by the strong-feedback variant of the FLAMINGO suite, with the highest-mass and highest-redshift bins hinting at even stronger feedback. Fitting the combined low-redshift \CM and \efeds gas fractions with a single mass-dependent BFC model, we derived constraints on the resulting suppression of the total matter power spectrum that are likewise consistent with a strong-feedback scenario, and that agree broadly, within their respective mass and scale sensitivities, with recent kSZ-based and X-ray-based constraints from the literature.

We further demonstrate that the mass-luminosity relation alone is not a reliable indicator of feedback strength. Although our measured $M$-$L_X$ relation, together with previous \efeds results, is compatible with a moderate (fiducial) feedback scenario, the corresponding gas fractions and power spectrum suppression point towards stronger feedback. We showed with a simple toy model that two density profiles sharing the same mass and luminosity, but differing in radial shape, can yield gas fractions consistent with either the fiducial or the strong-feedback FLAMINGO scenario, because luminosity scales as the square of the electron density while the gas fraction scales linearly with it. This degeneracy plausibly explains why our results and those of a recent independent analysis of groups in eRASS:1 data reach different conclusions on feedback strength despite agreeing on the scaling relation, and it illustrates more broadly why calibrating or constraining feedback from integrated quantities such as gas fractions, without reference to the underlying profile shape, can be misleading.

These findings motivate moving beyond gas-fraction-based calibration towards a direct, profile-level treatment of X-ray data. In a follow-up paper, we will apply the model developed here to fit the \CM and \efeds surface brightness and luminosity profiles directly, which is expected to remove much of the shape-driven degeneracy identified above and to yield correspondingly tighter and more robust constraints on baryonic feedback. Looking further ahead, the same framework can be used to forward-model the diffuse X-ray emission at the map level, opening a path towards simulation-based inference applied directly to wide-field X-ray surveys such as eROSITA and its successors.

\begin{acknowledgments}

The authors would like to thank Dominique Eckert, Riccardo Seppi, Emre Bahar, Scott Kay and Tilman Tröster for discussions and advice on various aspects of the project. We would also like to thank David Alonso for providing initial code and many helpful discussions. This work was supported in part by grant 20FL20\_201479 from the Swiss National Science Foundation. AN acknowledges support from the European Research Council (ERC) under the European Union’s Horizon 2020 research and innovation programme with Grant agreement No. 101163128. SKG acknowledges support from the Olle Engkvist Stiftelse (grant no. 232-0238). NC is supported by CNES. AL is supported by the National Natural Science Foundation of China, Grant No. 12588202. We acknowledge the computational resources of the Euler cluster at ETH Zurich and Claude Code (Anthropic) for assistance with code development and text refinement.
\end{acknowledgments}

\appendix
\section{Effects of temperature and metallicity modelling}
\label{app:T_modelling}

\begin{figure}
    \centering
    \includegraphics[width=1.\linewidth]{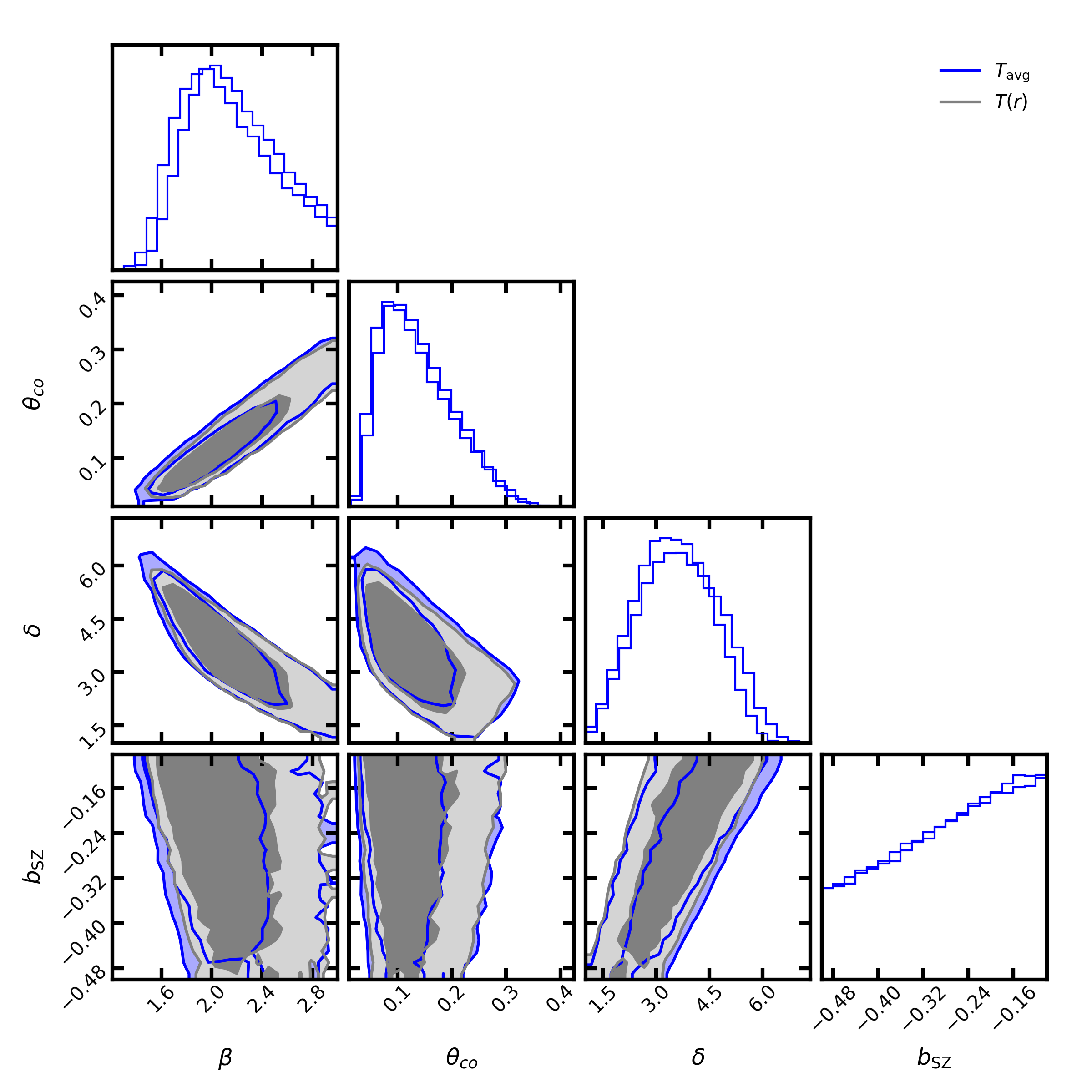}
    \caption{Posterior shift due to different modelling of the temperature profile. The blue set of contours is obtained by modelling the gas emissivity inside the clusters assuming the average gas temperature. Grey contours are obtained by modelling the full radial dependence of the ICM temperature.}
    \label{fig:validation_Tprofile_contours}
\end{figure}

Our fiducial choice in modelling the stacked profiles is to assume the average temperature for the cluster $T_{\rm avg}$ from Eq.~\eqref{eq:Tavg}, following numerous X-ray studies \citep{Bartalucci_2023, Lyskova_2023}. This approximation is motivated by a weak temperature dependence of the X-ray emissivity in massive clusters. In Fig.~\ref{fig:validation_Tprofile_contours} we show the effect of this approximation at the level of the MCMC contours. We show the marginalised posteriors from the fit to the \ls sample in blue. In the follow-up analysis, we replace the average temperature of a given cluster with the radial temperature profile $T(r)$ that can be computed using Eq.~\eqref{eq:T_BFC} and show the resulting projected posteriors as grey contours in the same figure. We observe a negligible difference between the two sets of contours, confirming the approximation does not introduce biases to our results. We also show the best-fit to the stacked profile and the corresponding electron number density profile as solid grey lines in Fig.~\ref{fig:validation_ne_lyskova}.

\begin{figure}
    \centering
    \includegraphics[width=1.\linewidth]{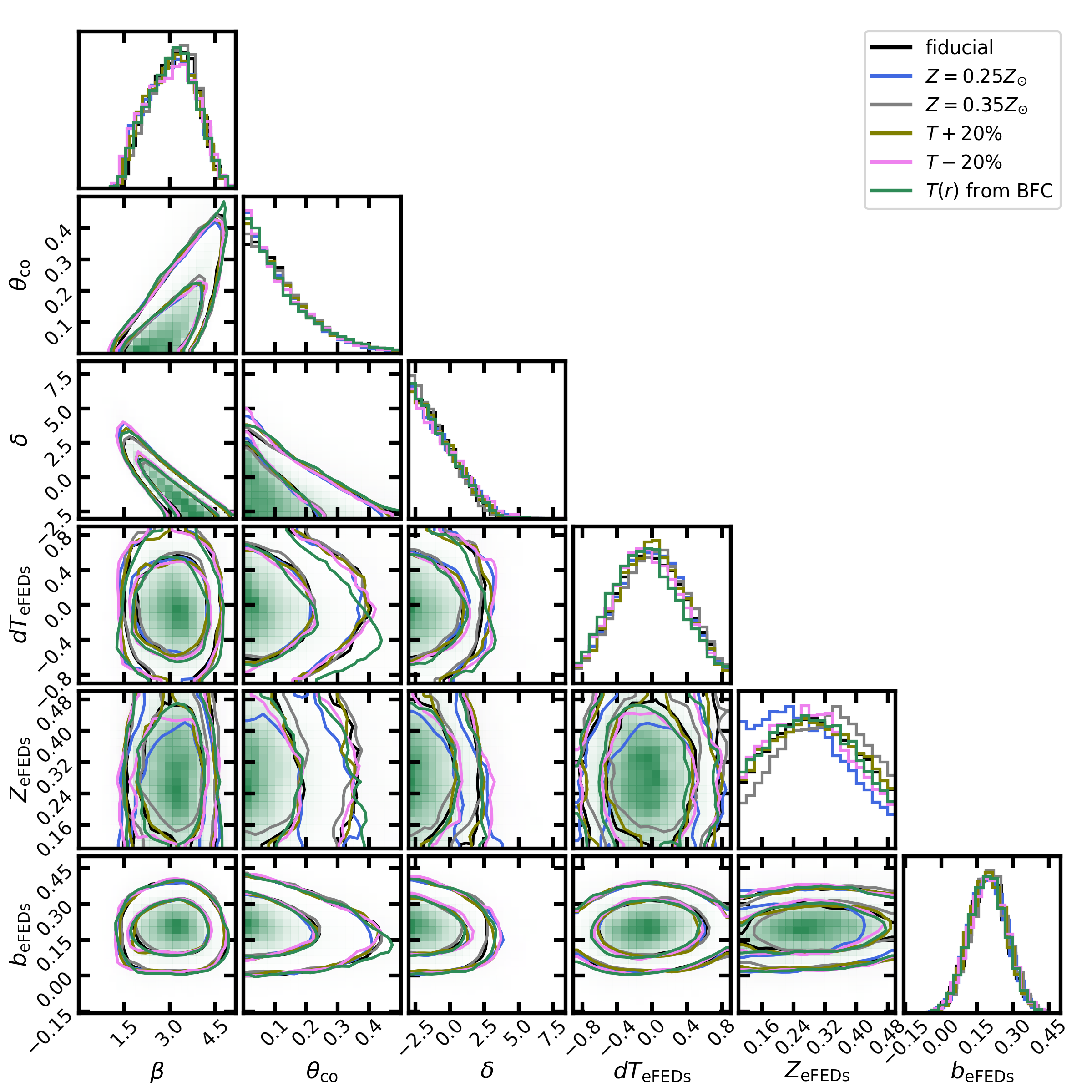}
    \caption{Posterior contours for \efeds bin~1 MCMC chains, varying temperature and metallicity modelling choices.}
    \label{fig:sensitivity_efeds}
\end{figure}

\begin{figure}
    \centering
    \includegraphics[width=1.0\linewidth]{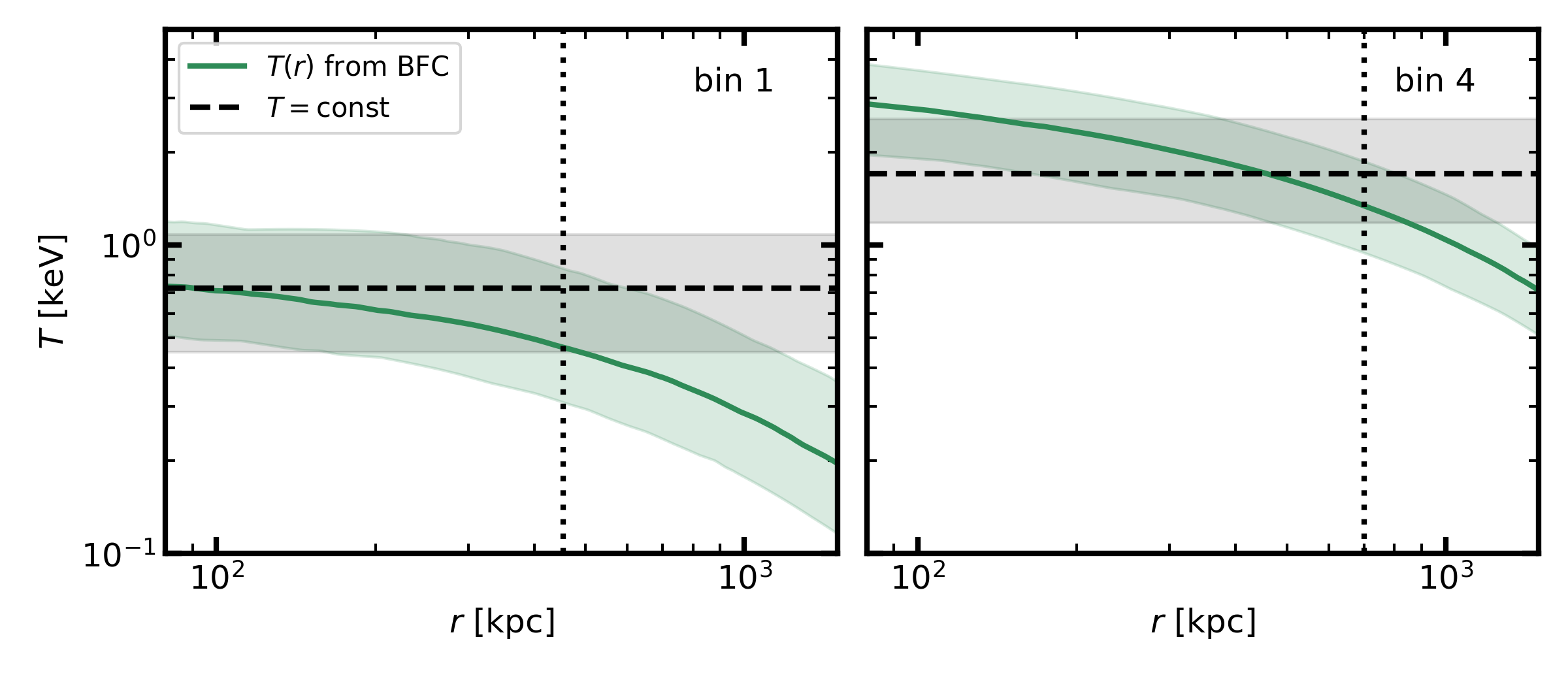}
    \caption{The radial temperature profiles (green) and constant temperatures (grey) as sampled in the MCMC chains, assuming the BFC radial temperature profile normalised to the temperature from the scaling relation of \cite{efeds_mass_calibration} and constant temperature from the same scaling relation.}
    \label{fig:Tprof_vs_Tconst_efeds}
\end{figure}

\section{Effects of mass bias}
\label{app:bias_effects}




\begin{figure}
    \centering
    \includegraphics[width=1.0\linewidth]{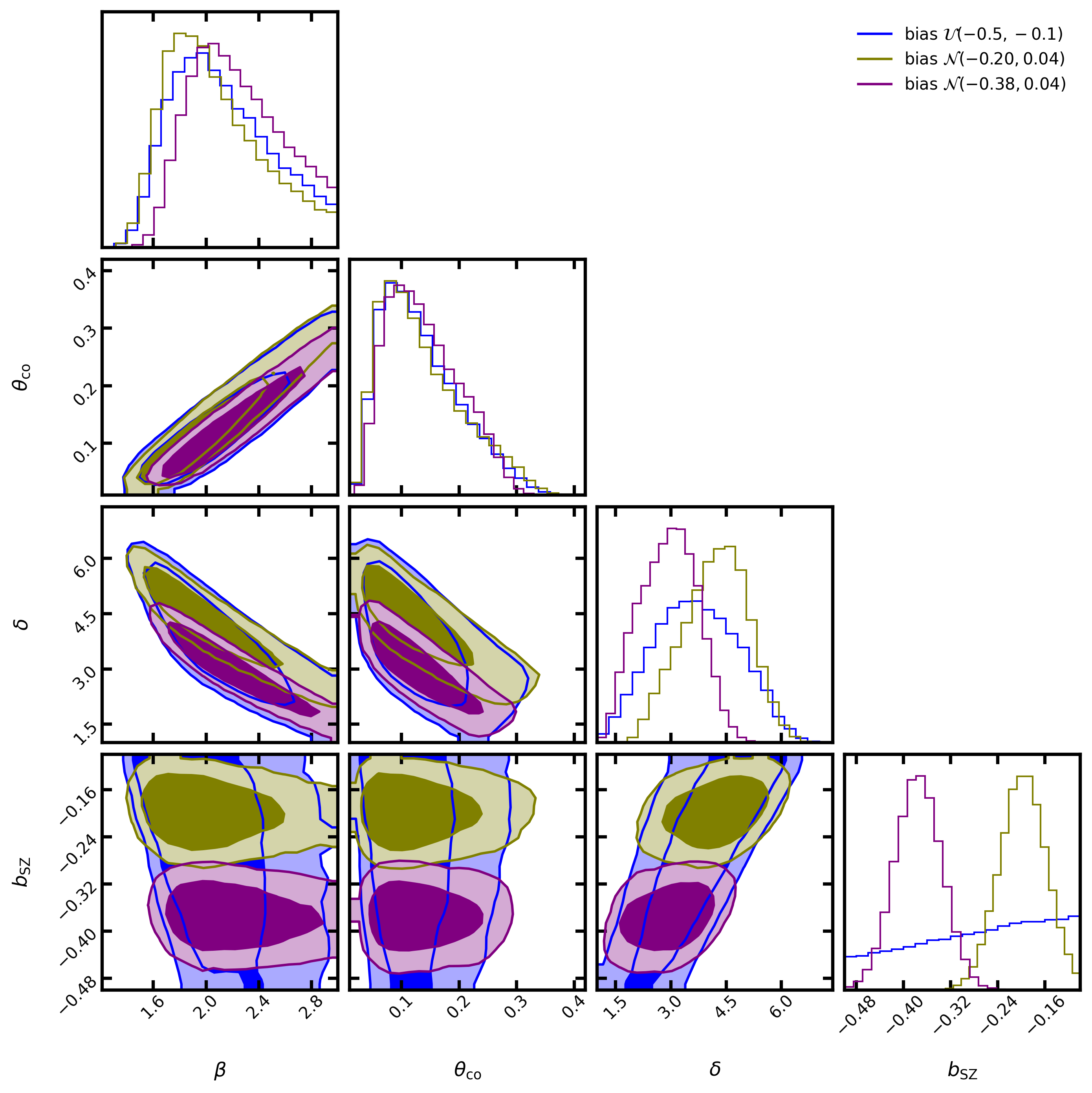}
    \caption{Impact of different bias modelling on the results assuming the median profile of the 38-cluster subsample of CHEX-MATE as presented in \cite{Lyskova_2023}. Blue contours are obtained assuming a flat prior on the bias parameter, while purple and olive correspond to scenarios with a Gaussian prior with mean $-0.38$ and $-0.20$, respectively.}
    \label{fig:mcmc_bin7_bias}
\end{figure}

In Fig.~\ref{fig:mcmc_bin7_bias}, we showcase the effects of SZ-bias as defined in Eq.~\eqref{eq:bSZ} on the BFC posteriors. For this example, we work with the \ls bin of \CM sample. In blue, we show the marginalised posteriors of BFC parameters for our fiducial choice of \bSZ{} prior, a flat conservative prior of $[-0.5,-0.1]$. This results in degeneracies of mass bias parameter and the two slope parameters $\beta$ and $\delta$ (revisit Eq.~\eqref{eq:rho_hga_def}), especially with $\delta$. This degeneracy can be understood as follows: with larger $\delta$-values, the hot gas profile is steeper at large radii, increasing its amplitude because the profile's normalisation remains the same. This amplitude growth needs to be compensated by a smaller overall mass of the profile, in order to fit the same observed profile. Therefore, with increased $\delta$, the \bSZ{} parameter needs to become closer to zero (thus reducing the total mass of a cluster).

As purple and olive contours in the same figure, we show the BFC posteriors under the assumption of Gaussian prior on the mass bias parameter, specifically for $\bSZ = -0.38 \pm 0.04$ \cite{Sereno_2025} and a less extreme value of $\bSZ = -0.20 \pm 0.04$, respectively. This leads to the degeneracy breaking between \bSZ{} and $\delta$; thus the latter shape parameter becomes constrained. Not surprisingly, these two choices of informative priors on mass bias will significantly affect the recovered gas fractions. We therefore opt for a more conservative prior on the mass bias in our main analysis.

\section{Forward modelling of \efeds sample and detection probabilities of its objects}
\label{app:forward_model}

The \efeds catalogue contains X-ray sources detected by eROSITA across a
$140\,\mathrm{deg}^2$ field. This field of view is very narrow even compared to the later eROSITA data releases. The resulting area-limited sampling of objects within a given mass and redshift range in \efeds observations, combined with the X-ray selection adopted by the collaboration, introduces selection biases. The selection functions depending on various observational quantities have been constructed for the \efeds sample, especially the one used by this work depends on $P_{\rm det}(L_X, z, t_{\rm exp}, \EM{0})$ (see Sec.~\ref{sec:data}) for more details), capturing the shape- and amplitude-based biases close to the detection limits of the eROSITA instrument. However, some objects in the sample receive extremely high detection probabilities, for example, faint objects with short exposure times. However, if a single object is a significant outlier in terms of its detection weight, it can completely dominate the statistical properties of a given \efeds mass-redshift bin. This can be a physical object (we happen to observe one of many faint objects that should be properly weighted in our binning procedure) or an error in the measured properties of such an object that results in an unphysically large detection probability.  In this section, we develop a simple forward model to distinguish these two scenarios.

In the first step, we build \efeds-like mock by sampling from a Tinker80 halo mass function implemented in \texttt{pyccl}\footnote{\url{https://ccl.readthedocs.io/en/latest/}}  over the comoving volume of the eFEDS footprint
($\Omega=140\,\mathrm{deg}^2$, $z\in[0.01,1.3]$,
$M_{500c}\in[5\times10^{12},10^{15}]\,M_\odot$). We adopt a flat $\Lambda$CDM cosmology with $\Omega_m=0.30$, $\Omega_b=0.049$, $h=0.70$, $n_s=0.96$, $\sigma_8=0.80$.
Each halo is assigned an X-ray luminosity drawn from the Chiu et al.\
(2022) $L_X$--$M_{500c}$--$z$ scaling relation \cite{efeds_mass_calibration} as described in the main text. A halo enters the mock catalogue only if it satisfies
$P_{\rm det}(L_X,z,t_{\rm exp})>u$, where $u$ is drawn uniformly from
$[0,1]$ and $t_{\rm exp}$ is drawn from an empirical two-component
log-normal mixture fitted to the real \efeds sample.
The observed mass is then obtained by adding weak-lensing scatter,
$M_{\rm obs}=M_{\rm true}\exp[\mathcal{N}(0,\sigma_M^2)]$ with
$\sigma_M=0.25$.
Finally, clusters are placed into the same eight $(M_{\rm obs},z)$ bins
used for the real data, and a KDE-based weight-pruning step is applied
within each bin so that the joint $(M_{\rm obs},z)$ distribution of the
mock matches that of the data before bin summaries are computed.

\begin{figure}
    \centering
    \includegraphics[width=1.\linewidth]{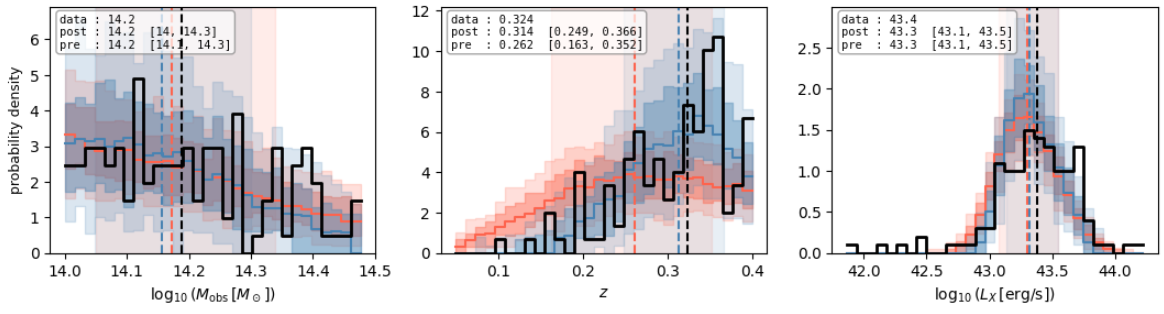}
    \caption{Illustration of \efeds forward model performance for bin~4. We show the distribution of observed masses (left), redshifts (middle) and virial luminosities (right) of the data (black) and median (solid line) and 16\%-84\% percentile range over the 100 mock \efeds realisations (in colour), red showing the model prediction before and blue after KDE pruning of the simulated \efeds catalogue. Dashed vertical lines and bands show the 16,50,84 percentiles of the data (black) and averaged models (red, blue). }
    \label{fig:efeds_fm_bin4}
\end{figure}

We assess the fidelity of the forward model by running 100 independent
realisations, one per random seed, and comparing three summary statistics
per bin --- median $M_{500c}$, median $z$, and median $L_X$ --- to their
observed values.
For each bin $b$ we compute the Hartlap-corrected Mahalanobis distance
\begin{equation}
  d^2_b
  = \bigl(\bm{s}_{\rm data}-\bar{\bm{s}}_{\rm sim}\bigr)^{\!\top}
    \bm{C}_{\rm H}^{-1}\,
    \bigl(\bm{s}_{\rm data}-\bar{\bm{s}}_{\rm sim}\bigr),
  \quad
  \bm{C}_{\rm H} = \frac{n_{\rm sim}-n_{\rm summ}-2}{n_{\rm sim}-1}\,\bm{C}_{\rm sim},
\end{equation}
where $\bar{\bm{s}}_{\rm sim}$ and $\bm{C}_{\rm sim}$ are the mean and
sample covariance across the 100 mocks, and the Hartlap prefactor
corrects for noise bias in the precision matrix.
Under the null hypothesis $d^2_b$ follows a $\chi^2(3)$ distribution;
the corresponding p-value $p_{\chi^2}$ is reported in
Table~\ref{tab:fidelity}. We see that except for bin~5 (a sparse high-mass low-redshift bin, which is unbiased and does not contain any objects with spurious values of the selection function), the observed \efeds bins fall well within the distribution of 100 forward-modelled mocks ($p>0.05$). Fig.~\ref{fig:efeds_fm_bin4} illustrates the performance of the \efeds\ forward model in mass bin~4, comparing the observed cluster sample (black) to $100$ mock \efeds\ realisations drawn from the forward model (colour). For each of the three observables shown --- mass (left), redshift (middle), and virial luminosity (right) --- we show the median (solid line) and $16\%$--$84\%$ percentile range across the mock realisations, both before (red) and after (blue) KDE-based pruning of the simulated catalogue. The vertical dashed lines and shaded bands mark the $16$th, $50$th, and $84$th percentiles of the data (black) and of the corresponding averaged model distributions (red, blue).

Each detected cluster is assigned an inverse-probability weight
$w_i = 1/P_{\rm det}(L_X^{(i)}, z^{(i)}, t_{\rm exp}^{(i)})$,
which up-weights clusters near the detection threshold to correct the
flux-limited sample for selection bias.
A cluster with a very large weight was barely detectable: either it is a
genuine but rare extreme cluster, or it is a noise fluctuation that
passed the detection threshold by chance.
To remove the latter, we set a data-driven weight threshold per bin.

To model the tail of the weight distribution robustly, we generate a
large-statistics mock (two times the full sky) and fit a smoothly broken
power law (SBPL) to the histogram of $\log_{10}(w)$ in each bin.
The SBPL has three slopes and two break positions, and is fitted by
minimising a weighted chi-squared.

From the fitted SBPL we derive the expected distribution of the largest
weight in a sample of exactly $N_{\rm \efeds}$ clusters via the
max-order-statistic,
\begin{equation}
  p(w_{\rm max}) = N_{\rm \efeds}\,
    F(w_{\rm max})^{N_{\rm \efeds}-1}\,f(w_{\rm max}),
\end{equation}
where $f$ and $F$ are the SBPL density and CDF.
We then set the weight threshold $w_{\rm cut}$ to the 95th percentile
of this distribution: a cluster whose weight exceeds $w_{\rm cut}$ would
be surprising even in a noise-free realisation of the \efeds survey, and
is therefore flagged as a candidate spurious detection.
All flagged clusters are removed from the data vector before the
profile-fitting analysis. After this procedure, our final sample consists of 490 objects.

\begin{table}
  \centering
  \caption{Forward-model fidelity per bin. $p_{\chi^2}$ is the
           $\chi^2(3)$ p-value for the Hartlap-corrected Mahalanobis
           distance between the data and the mock ensemble.}
  \label{tab:fidelity}
  \begin{tabular}{c|cccccccc}
    \hline\hline
    Bin & 1 & 2 & 3 & 4 & 5 & 6 & 7 & 8 \\
    \hline
    $p_{\chi^2}$ & 0.461 & 0.332 & 0.084 & 0.083 & 0.003 & 0.533 & 0.207 & 0.206 \\
    \hline
  \end{tabular}
\end{table}

\section{MCMC posteriors from gas fraction analysis}
\label{app:mcmc_posteriors}

\begin{figure}
    \centering
    \includegraphics[width=1.\linewidth]{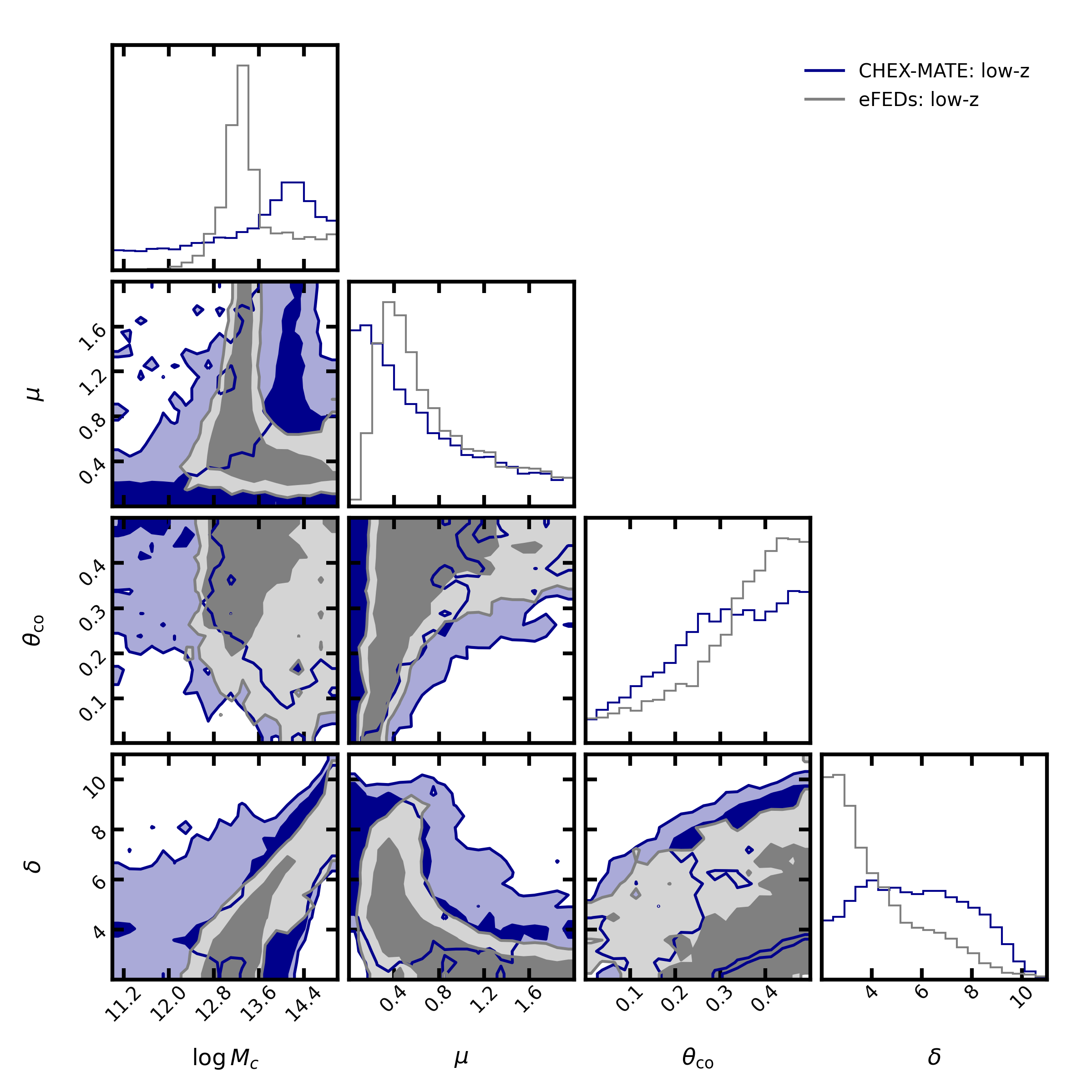}
    \caption{Posterior distributions of BFC parameters from the joint fit to low-z \CM (dark blue) and \efeds (grey) gas fractions. Shown are 68\% and 95\% C.I.}
    \label{fig:fgas_contours}
\end{figure}

In Fig.~\ref{fig:fgas_contours} we show the marginalised posteriors of the fit to the low-z gas fraction measurements of \CM and \efeds, discussed in Sec.~\ref{sec:results}. We only show the four shared BFC parameters between the two runs. The dark blue contours show the posteriors of \CM and the black contours those of the \efeds sample. The contours are generally in good agreement and are shown at 68\% and 95\% C.I. 

\bibliographystyle{JHEP}
\bibliography{main}

\end{document}